\documentclass[useAMS,usenatbib]{mn2e}
\usepackage{txfonts}
\usepackage{graphicx}
\usepackage{aas_macros}

\newcommand{\sext}{{\sl SExtractor}}

\newcommand{\lya}{Ly$\alpha$}
\newcommand{\ha}{\ifmmode{\mathrm{H\alpha}}\else{H$\alpha$}\fi}
\newcommand{\hb}{\ifmmode{\mathrm{H\beta}}\else{H$\beta$}\fi}
\newcommand{\hc}{\ifmmode{\mathrm{H\gamma}}\else{H$\gamma$}\fi}
\newcommand{\Ha}{\ha~$\lambda$6564}
\newcommand{\Hb}{\hb~$\lambda$4863}
\newcommand{\Hc}{\hc~$\lambda$4342}
\newcommand{\oi}{[{\sc Oi}]}
\newcommand{\oii}{[{\sc Oii}]}
\newcommand{\oiii}{[{\sc Oiii}]}
\newcommand{\Oi}{\oi~$\lambda$6302}
\newcommand{\Oii}{\oii~$\lambda\lambda$3726,3728}
\newcommand{\Oiii}{\oiii~$\lambda\lambda$4959,5007}
\newcommand{\Oiiib}{\oiii~$\lambda$5007}
\newcommand{\nii}{\ifmmode{\mbox{{[\sc N\,ii]}}}\else{{\sc [N\,ii]}}\fi}
\newcommand{\Nii}{\nii~$\lambda\lambda$6550,6585}

\newcommand{\Niib}{\nii~$\lambda$6585}
\newcommand{\sii}{{\sc [S\,ii]}}
\newcommand{\Sii}{\sii~$\lambda\lambda$6733,6718}

\newcommand{\HII}{{\sc Hii}}

\newcommand{\Mpc}{Mpc$^3$}
\newcommand{\perMpc}{Mpc$^{-3}$}
\newcommand{\perMpcSq}{Mpc$^{-2}$}

\newcommand{\kmsMpc}{km\,s$^{-1}$\,Mpc$^{-1}$}
\newcommand{\ergs}{\ifmmode{\mathrm{erg\,s^{-1}}}\else{erg\,s$^{-1}$}\fi}

\newcommand{\lineunits}{erg\,s$^{-1}$\,cm$^{-2}$}

\newcommand{\Msunyr}{M$_\odot$\,yr$^{-1}$}

\newcommand{\fluxunits}{\lineunits}

\newcommand{\arcdeg}{\mbox{$^\circ$}}

\renewcommand{\d}{\arcdeg}

\newcommand{\as}{\arcsec}

\newcommand{\na}{$N_{810}$}
\newcommand{\nb}{$N_{817}$}
\newcommand{\nc}{$N_{824}$}

\providecommand{\pow}[2][10]{#1^{#2}}

\providecommand{\BRcol}{\ifmmode(B-R)\else$(B-R)$\fi}
\providecommand{\R}{$R$}
\providecommand{\B}{$B$}

\providecommand{\fwhm}{{\it FWHM}}

\newcommand{\eqref}[1]{(\ref{#1})}

\providecommand{\includeIDLfig}[2][\columnwidth]{\includegraphics[width=#1, trim=20pt 8pt 5pt 18pt, clip]{#2}}
\providecommand{\centercol}[1]{\multicolumn{1}{c}{#1}}
\begin{document}

\title[SFD and \ha{} LF of an Emission Line Selected Galaxy Sample at
$z\sim0.24$]{Star Formation Density and \ha{} Luminosity Function of
  an Emission Line Selected Galaxy Sample at $z\sim0.24$ \thanks{Based
    on observations made with ESO Telescopes at the La Silla
    Observatory (Programmes 67.A-0063, 68.A-0363 and 69.A-0314) and
    the Anglo-Australian Telescope.}}

\author[Westra \& Jones]{Eduard Westra$^1$\thanks{Email:
    westra@mso.anu.edu.au} and D. Heath Jones$^2$\thanks{Email:
    heath@aao.gov.au}\\
  $^1$Research School of Astronomy \& Astrophysics, The Australian
  National University, Cotter Road, Weston Creek ACT 2611, Australia\\
  $^2$Anglo-Australian Observatory, PO Box 296, Epping NSW 1710,
  Australia}

\date{Accepted 2007 October 3. Received 2007 September 12; in original
form 2007 May 28}

\maketitle

\label{pageFirst}

\begin{abstract}
  We use narrowband imaging (\fwhm{} $=70$\,\AA{}) to select a sample
  of emission line galaxies between $0.20 \lesssim z \lesssim 1.22$ in
  two fields covering 0.5 sq.~deg. We use spectroscopic follow-up to
  select a sub-sample of \ha{} emitting galaxies at $z\sim0.24$ and
  determine the \ha{} luminosity function and star formation density
  at $z\sim0.24$ for both of our fields. Corrections are made for
  imaging and spectroscopic incompleteness, extinction and interloper
  contamination on the basis of the spectroscopic data. When compared
  to each other, we find the field samples differ by $\Delta \alpha =
  0.2$ in faint end slope and $\Delta \log [ L^* (\mathrm{\ergs}) ] =
  0.2$ in luminosity. In the context of other recent surveys, our
  sample has comparable faint end slope, but a fainter $L^*$
  turn-over. We conclude that systematic uncertainties and differences
  in selection criteria remain the dominant sources of uncertainty
  between \ha{} luminosity functions at this redshift.

  We also investigate average star formation rates as a function of
  local environment and find typical values consistent with the field
  densities that we probe, in agreement with previous results.
  However, we find tentative evidence for an increase in star
  formation rate with respect to the local density of star forming
  galaxies, consistent with the scenario that galaxy-galaxy
  interactions are triggers for bursts of star formation.
\end{abstract}

\begin{keywords}
  surveys -- galaxies: luminosity function, mass function -- galaxies:
  starburst
\end{keywords}

\defcitealias{Westra06}{Paper I}

\section{Introduction}

It is now widely accepted that the amount of star formation in
Universe as a whole has increased since the formation of the first
galaxies, peaking around redshifts $z\sim2-3$ and subsequently
declining by a factor of ten \citep[e.g.][and references
therein]{Hopkins04}. Cosmic star formation history provides strong
constraints on models of galaxy formation and evolution
\citep{Pei99,Somerville01}, because it directly traces the
accumulation of stellar mass and metal fraction \citep{Pei95,Madau96}
to their present-day values \citep{Cole01,Panter03}. Its rapid decline
over the past 8 Gyr is consistent with ``downsizing'' scenarios
\citep{Cowie96} in which the more massive galaxies have produced their
stellar mass at earlier times than the less massive galaxies
\citep{Heavens04,Juneau05,Thomas05,Fardal06}. The star formation
history of the universe has also been used to constrain allowable
stellar initial mass functions \citep{Baldry03,Hopkins06} and cosmic
supernova rates \citep{GalYam04,Daigne06}.

Star forming galaxies exhibit a strong UV continuum courtesy of newly
formed OB stars in sites of star formation. This newborn population
can be inferred from the UV directly \citep[e.g.][]{Treyer98,Lilly96}
or through a host of indirect calibrators spread across the
electromagnetic spectrum \citep{RosaGonzalez02,Condon92,Schaerer00}.
At low redshifts the most direct calibrator -- and of the optical
calibrators the least affected by internal extinction -- is the \ha{}
recombination line, which emits when stimulated by ionising UV
radiation \citep[e.g.][]{Kennicutt98}.

Narrowband surveys at optical wavelengths have long been recognised as
a powerful way of yielding large samples of emission line galaxies,
including those selected by \ha{} at redshifts $z\lesssim0.4$
\citep{Ly07,Pascual07,Jones01}. They are advantageous in that they
select galaxies in exactly the same quantity that they seek to
measure, and are optimised for the detection of the faint emission
line signatures indicative of star formation. Narrowband surveys also
have the advantage of a simplified selection function, with filters
that probe only a very narrow redshift slice, thereby yielding a
volume limited sample at a common distance. Many recent emission line
surveys have targeted \lya{} at high redshift
\citep{Ajiki03,Hu04,Rhoads04,Gawiser06}, as well as \ha{}, \hb{},
\oiii{} and \oii{} at lower redshifts
\citep{Fujita03,Hippelein03,Ly07}.

Here we describe a survey for \ha{} emission line galaxies at
$z\sim0.24$, found as a by-product of the Wide Field Lyman Alpha
Search \citep[WFILAS;][]{Westra05,Westra06}. The resulting sample has
been utilised to determine the \ha{} luminosity function at
$z\sim0.24$ and its associated co-moving star formation density. In
Section~\ref{sec:candsel} we describe the selection of candidates
using narrow- and broadband imaging. In Section~\ref{sec:spectroscopy}
we detail follow-up spectroscopy used to identify the nature of the
emission and test completeness of the sample. In
Section~\ref{sec:lumfieSFD} we derive the \ha{} luminosity function
for galaxies at $z\sim0.24$ and explore its variation with the local
environment in Section~\ref{sec:environment}. A summary and concluding
remarks are made in Section~\ref{sec:conclusionsLowz}.

Throughout this paper we assume a flat Universe with $(\Omega_{\rm m},
\Omega_{\Lambda}) = (0.3,0.7)$ and a Hubble constant $H_0 =
70$\,\kmsMpc. All quoted magnitudes are in the {\it AB} system
\citep{Oke83}\footnote{$m_{AB} = -2.5 \log f_\nu - 48.590$, where
  $m_{AB}$ is the {\it AB} magnitude and $f_\nu$ is the flux density
  in ergs\,s$^{-1}$\,cm$^{-2}$\,Hz$^{-1}$}.

\section{Candidate selection}
\label{sec:candsel}

\subsection{Narrowband imaging}
\label{subsec:imaging}
The observations were done with the Wide Field Imager (WFI) on the
ESO/MPI 2.2\,m telescope at the Cerro La Silla Observatory, Chile. The
WFI consists of a four by two array of 2k\,$\times$\,4k CCDs giving a
total field size of 34\as{}\,$\times$\,33\as with pixel scale of
0\farcs238 per pixel. Imaging data were taken from the Wide Field
Lyman Alpha Search \citep[WFILAS;][]{Westra05,Westra06}, a wide-field
narrowband survey designed to find Lyman-$\alpha$ emitters at
$z\sim5.7$. We refer the reader to \citet[hereafter
\citetalias{Westra06}]{Westra06} for a more detailed description, but
give the important features of the survey below.

\begin{table*}
  \centering
  \begin{tabular}{cccccc}
    \hline\hline
     & \multicolumn{5}{c}{Emission line}\\
     & \multicolumn{1}{c}{\ha{}} & \multicolumn{1}{c}{\hb{}} &
    \multicolumn{1}{c}{\oiii{}} & \multicolumn{1}{c}{\oii{}} & \multicolumn{1}{c}{\sii{}}\\
    \hline
    Redshift range in \na{} & 0.229 -- 0.239 & 0.659 -- 0.673 & 0.610 -- 0.624 & 1.163 -- 1.182 & 0.199 -- 0.210\\
    Redshift range in \nb{} & 0.239 -- 0.250 & 0.673 -- 0.687 & 0.624 -- 0.638 & 1.182 -- 1.201 & 0.210 -- 0.220\\
    Redshift range in \nc{} & 0.250 -- 0.261 & 0.687 -- 0.702 & 0.638 -- 0.652 & 1.201 -- 1.219 & 0.220 -- 0.230\\
    \\
    $D_L$ (Mpc) & 1203.1 & 4081.9 & 3726.3 & 8158.4 & 1045.1\\
    $V_\mathrm{CDFS}$ ($\pow{3}$\,\Mpc{}) & 9.4 & 60.6 & 53.6 & 137.2 & 7.3\\
    $V_\mathrm{S11}$ ($\pow{3}$\,\Mpc{}) & 8.3 & 53.2 & 47.1 & 120.6 & 6.4\\
    \hline\hline
  \end{tabular}

  \caption{Redshift coverage, luminosity distance $D_L$, and co-moving
    volume for each emission line in each of our narrowband
    filters \na{}, \nb{} and \nc{} using the central wavelength
    and \fwhm{} of each filter (70\,\AA{}). The CDFS and S11 fields
    span differing volumes ($V_\mathrm{CDFS}$ and $V_\mathrm{S11}$,
    respectively). For \oiii{} we used the wavelength of the
    \oiii{}~$\lambda$\,5007 line and for \oii{} and \sii{} the average
    wavelength of the individual lines of each doublet.}
  \label{tab:filterredshifts}
\end{table*}

Three fields spaced around the sky were observed in three narrowband
filters (\fwhm{}\,=\,7\,nm) centred at 810, 817 and 824\,nm, an
intermediate width filter (\fwhm{}\,=\,22\,nm) centred at 815\,nm and
broadbands \B{} and \R{}. For one of the fields with missing 817\,nm
data it was not possible to apply the selection criteria uniformly and
so it was excluded from this analysis. The two fields used were the
well-studied Chandra Deep Field South \citep[CDFS;
e.g.][]{Rosati02,Rix04} and the COMBO-17 S11 field \citep{Wolf03}. The
width of our narrowband filters is essentially half that of other
surveys \citep[e.g.][]{Fujita03,Ly07} with a corresponding reduction
in background and enhancement in the contrast of observations of
emission line galaxies. Table~\ref{tab:filterredshifts} gives an
overview of the emission lines redshifted into these narrowband
filters, the associated luminosity distances and co-moving volumes.

The data were processed using a combination of standard
IRAF\footnote{IRAF is distributed by the National Optical Astronomy
  Observatories, which are operated by the Association of Universities
  for Research in Astronomy, Inc., under cooperative agreement with
  the National Science Foundation.} routines ({\tt mscred}) and some
custom designed for our data. Image frames were bias-subtracted,
flat-fielded and background-subtracted. A fringe pattern present in
the intermediate band and narrowband images, which remained after the
flat-fielding, was removed using a fringe frame created from 10--30
science frames. Finally, an astrometric correction was applied using
the USNO CCD Astrograph Catalogue 2 \citep[UCAC2;][]{Zacharias04} and
the IRAF-task {\tt msccmatch} with a resulting RMS of $\lesssim
0\farcs15$.

To ensure the quality of the final deep images we only included frames
with a seeing of less than 5 pixels (=1\farcs2) and without
significant fringing. The images were weighted according to their
exposure time and combined using the IRAF {\tt mscstack} routine
rejecting deviant pixels.

\subsection{Photometry and completeness corrections}
\label{sec:photcompcorr}
We used \sext{} \citep[version 2.3.2; ][]{Bertin96} in double image
mode to do create the initial source catalogues. Each resulting
catalogue contains the photometry for the sources in all 6 filters.
Sources were selected when at least 5 pixels were 0.8\,$\sigma$ above
the noise level in the narrowband image used for detection. All
photometry was measured in apertures with a 10 pixel diameter (=
2\farcs4). \citetalias{Westra06} describes the procedure in detail.

Detection completeness was determined using galaxy number-counts in
each of the narrowband images as a function of {\it AB}-magnitude and
that of the Hubble Deep Field (HDF) in the {\it F}814{\it W} filter
\citep{Williams96}. Completeness is defined in this instance as the
ratio of the number of detected galaxies to that of expected, and the
completeness correction is its reciprocal. The expected number counts
were fit by a simple linear function over the magnitude range [20,
25]. For all the objects that are selected as our candidates this
correction is less than 0.1\,\%.

\subsection{Selection criteria and star/galaxy disambiguation}
\label{sec:selectionCritera}
The following four criteria were applied to select our candidate
emission-line galaxies from the initial source catalogues:
\begin{enumerate}
\item the narrowband image used as the detection image must have the
  most flux of all the narrowband images and the source must have a
  4\,$\sigma$ detection or better in the detected narrowband;
\item there must be at least a 2\,$\sigma$ detection in the
  intermediate band image;
\item \label{critBB} the broadband image \R{} needs to have a
  2\,$\sigma$ detection or better;
\item \label{critFlux} the emission line flux calculated from the
  narrowband images should be
  $F_\mathrm{line}\,\geq$\,$\pow{-16}$\,\fluxunits{}.
\end{enumerate}

Criterion~\ref{critBB} removes emission line objects with a very low
continuum. These objects were classified as \lya{} emitters at
$z\sim5.7$. The \lya{} emitters are discussed in
\citetalias{Westra06}. The emission line fluxes that we use in this
paper were measured from the narrowband photometry. The background (or
underlying continuum) was determined by averaging the flux measured in
the two narrowband images that were \textit{not} used for the
detection of the source. This was subtracted from the flux measured in
the narrowband detection image, which is emission line and continuum
flux combined. An aperture correction was calculated according to:
\begin{equation}
  C = \mathrm{max}(0.2, \mathrm{erf}\left (\frac{10}{2 a} \right )
  \times \mathrm{erf}\left (\frac{10}{2 b}\right ))
\end{equation}
and applied to the line fluxes. Here, $C$ is the fraction of light of
the object contained within the 10 pixel aperture, $a$ and $b$ are the
profile width along the major and minor axes, respectively (assuming
that the galaxy profile is adequately represented by a two-dimensional
Gaussian) and $\mathrm{erf}(x)$ is the error function\footnote{The
  error function is defined as
  $\mathrm{erf}(x)\,=\,\frac{2}{\sqrt{\pi}} \int \limits _0 ^x
  e^{-t^2} dt$}. To ensure that the fluxes of certain large objects
were not over-corrected, we limited $C$ to at least 0.2. Dividing the
calculated emission flux by $C$ gives the emission line flux
$F_\mathrm{line}$ used in criterion~\ref{critFlux}.

The emission line flux limit of $\pow{-16}$\,\fluxunits{} is a factor
of two higher than the detection limit of our earlier search for high
redshift \lya{} emitting galaxies using the same imaging data
($F_\mathrm{limit}$\,=\,5\,$\times$\,$\pow{-17}$\,\fluxunits;
\citetalias{Westra06}). This is because we are no longer limited by
the night-sky background, but rather by the brightness of the object
continua. This limit was chosen in part to ensure that emission line
candidates were within the sensitivity limits of our follow-up
confirmation spectroscopy. We note that we used a flux limit rather
than an equivalent width cut-off. The lowest equivalent width values
as determined from the narrowband imaging in our candidate sample is
$\sim1$\,\AA{}, with a peak at $\sim3$\,\AA{}.

\begin{figure}
  \centering
  \includeIDLfig{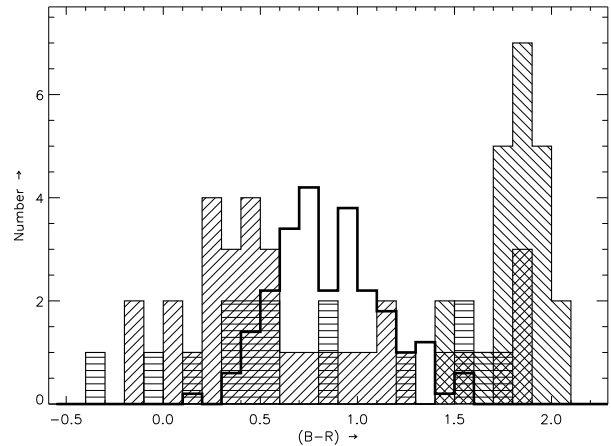}
  \caption{Distribution of observed \BRcol{} colour for narrowband
    candidates satisfying various criteria for stellarity: ({\it a})
    \sext{} {\sc class\_star}~$\ge$~0.95 and $R<21$ ({\it forward
      cross-hatching}), ({\it b}) \sext{}
    $\frac{\textsc{a\_image}}{\textsc{b\_image}}$\,$\le$\,1.06 and
    $\BRcol{}\ge1.4$ ({\it backward cross-hatching}) and ({\it c}) bright
    stars showing diffraction spikes or ghost reflections ({\it
      horizontal cross-hatching}). The thick histogram shows the
    combined distribution of \ha{} and \sii{} galaxies ($z\sim 0.24$
    and $z\sim0.21$, respectively) from our full emission line sample
    subsequently through follow-up spectroscopy
    (Section~\ref{subsec:spectroscopy}), scaled by 0.2.}
  \label{fig:BRcol_histo_stars}
\end{figure}

Stars represent a significant fraction of contaminants. We found that
standard star/galaxy classification from \sext{} works satisfactorily
for objects brighter than $R=21$. However, it breaks down for the
large number of faint ($R>21$) objects. Therefore, additional criteria
were applied. We examined the size of the objects (major and minor
axes), in combination with their shape (the ratio of the major and
minor axes) as additional star/galaxy discriminants. Since this
size/shape information could potentially lead to the unwanted removal
of unresolved line emitting galaxies, we used an additional cut in
\BRcol{} colour as a safeguard to prevent this\footnote{We note that
  this colour selection will also reject QSOs. This is of no
  consequence to our selection of a star forming sample.}. We decided
to restrict the size/shape discrimination to sources with
\BRcol{}\,$\ge$\,1.4 based on the \BRcol{} colour distribution of
\ha{} emitters at $z\sim0.24$ and \sii{} emitters at $z\sim0.21$
obtained from spectroscopic observations. This size/shape/colour
criterion was added after initial spectroscopic follow-up to improve
removal of stellar contaminants. Furthermore, the colour of $\BRcol{}
= 1.4$ corresponds to the model of an instantaneous star-burst with an
age of $\sim1$\,Gyr. We determined this colour using GALAXEV
\citep{Bruzual03}. We finalised our stellar selection criteria as
follows:
\begin{enumerate}
\item \label{it:sex} the \sext{} {\sc class\_star} parameter is
  $\ge$\,0.95 and $R$\,$<$\,21. At $R$\,$>$\,21, sources are too faint
  for \sext{} to reliably distinguish between stars and galaxies;
\item \label{it:idl} the \sext{} {\sc a\_image} and {\sc b\_image}
  parameters (the profile in pixels along the major and minor axes,
  respectively) are $\le$\,4\,pixels, the ratio of these parameters is
  $\frac{\textsc{a\_image}}{\textsc{b\_image}}$\,$\le$\,1.06 and the
  object has a \BRcol{} colour $\ge$\,1.4. This is redder than almost
  all star forming galaxies at $z \sim 0.24$;
\item \label{it:man} the object showed obvious imaging artefacts,
  such as diffraction spikes or ghost images, in any of its
  thumbnails.
\end{enumerate}
Figure~\ref{fig:BRcol_histo_stars} shows the distribution of
spectroscopically observed objects that satisfy these criteria as a
function of observed \BRcol{} colour for the CDFS field. The forward
cross-hatched histograms represent objects satisfying
criterion~\ref{it:sex}, the backward cross-hatched those for
criterion~\ref{it:idl}, and the horizontal cross-hatched those for
criterion~\ref{it:man}. The histogram outlined by the thick solid line
represents the observed \BRcol{} colour distribution of securely
confirmed \ha{} and \sii{} emitters ($z\sim0.24$ and $z\sim0.21$,
respectively), by way of comparison. All objects selected in this way
were deemed to be stellar and removed from the candidate list.
Finally, all candidates were inspected to remove sources that were
contaminated by image artefacts.

From initial candidate numbers of 786 and 848 for the CDFS and S11
fields respectively, 414 and 513 candidates were removed because they
met one or more of the stellar criteria. Our final sample yielded 372
candidate emission-line galaxies for the CDFS field and 335 for the S11 field.

\section{Spectroscopic follow-up}
\label{sec:spectroscopy}

\subsection{Observations and reduction}
\label{subsec:spectroscopy}
\begin{figure}
  \centering
  \includeIDLfig{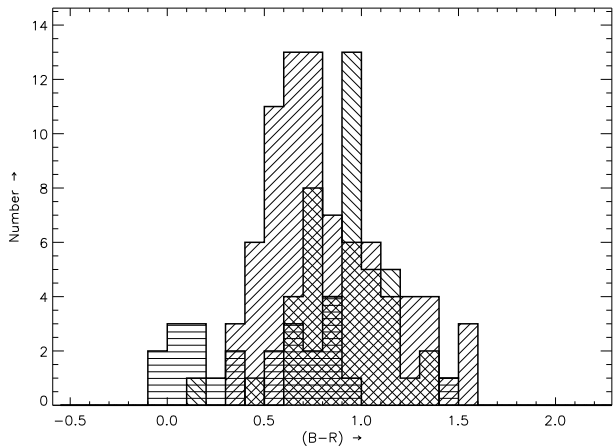}
  \caption{Observed \BRcol{} colour distribution for various sets of
    emission-line galaxies within our sample: ({\it a}) \ha{} at
    $z\sim0.24$ ({\it forward cross-hatching}), ({\it b}) \sii{} at
    $z\sim0.21$ ({\it backward cross-hatching}) and ({\it c})
    single-line emitters of indeterminate origin ({\it horizontal
      cross-hatching}).}
  \label{fig:BRcol_histo_elg}
\end{figure}

The emission-line selection criteria established in
Section~\ref{sec:selectionCritera} are sensitive to almost any galaxy
with emission lines that have been redshifted into the wavelength
range of our narrowband filters, and are bright enough to be detected.
The one exception is \lya{}, which does not yield detectable flux
blueward of the Lyman limit and hence in our broadband images. The
main emission lines to expect in our narrowband filters are (from
bluest to reddest), \Oii{}, \Hc{} (although usually too faint, or too
much underlying absorption), \Hb{}, \Oiii, \Ha{} and \Sii{}. Since the
goal of this paper is to establish the star formation density at
$z\sim0.24$, we concentrated only on those galaxies detected as \ha{}.
Alternative approaches by other groups \citep[e.g.][]{Ly07} have
separated objects based on their broadband colours. Unfortunately, in
the case of \sii{} galaxies ($z\sim0.21$) the colours are
indistinguishable from those with \ha{} ($z\sim0.24$) due to their
similar redshifts. Figure~\ref{fig:BRcol_histo_elg} shows how the
\ha{} and \sii{} galaxies occupy the same range of colour [$\BRcol{}
\gtrsim 0.5$] given their near-identical redshifts. Based on this, we
classify all of the single-line emitters outside this range [$\BRcol{}
\leq 0.5$] as likely \oii{} line-emitters at $z\sim1.2$. It is worth
pointing out that when the \sii{} doublet falls inside our narrowband
filter set, an extra volume of about 50\,\% of the volume probed by
\ha{} can be explored. Unfortunately, the fluxes of \sii{} and \ha{}
are not sufficiently correlated to permit star formation density
determinations from the \sii{} line \citep[e.g.][]{Kewley01}, and so it
was not used.

Our approach was to target as large a sample as possible of our
candidates to test how successful our candidate selection was. An
additional aim was to measure the fraction of the observed candidates
with \ha{} in our narrowband filters. To do this, we ensured that the
spectroscopic sample was representative of the narrowband sample as a
whole. A two-sided Kolmogorov-Smirnov test yielded probability levels
of 99.8\,\% and 49.3\,\% for the CDFS and S11 fields, respectively.
Once measured, we applied the determined fraction to our entire sample
of candidates in each field.

The spectroscopic data were taken with AAOmega \citep{Sharp06}, an
optical multi-object spectrograph. It is fibre-fed from the prime
focus of the Anglo-Australian Telescope (AAT) by the 2dF facility
\citep{Lewis02a} to a dual-beam spectrograph, which in our case was
used with spectral ranges 3800--5700\,\AA{} and 5700--8700\,\AA{}. The
resolving power was $\delta \lambda = 3.5$\,\AA{} in the blue arm and
$\delta \lambda = 5.3$\,\AA{} in the red arm. It has 392 fibres
available to observe spectra of objects within a 2 degree field of
view. The fibres have a minimum placement separation of 30\arcsec{},
although the actual limiting separation depends on the orientation of
fibre buttons when placed on the field plates. For fields with a high
density of targets, such as our 0.5\d{}\,$\times$\,0.5\d{} fields,
only $\sim$\,250 fibres could be allocated per configuration, due to
such placement limitations. In general, the number of fibres allocated
depends upon the target distribution in the field and the choice of
algorithm in the {\tt configure}\footnote{{\tt configure} and {\tt
    drcontrol} are software packages produced and maintained by the
  AAO. These packages can be obtained from {\tt
    ftp://ftp.aao.gov.au/pub/2df}} software. We found that using the
Simulated Annealing algorithm \citep{Miszalski06} allowed a larger
fraction of fibres to be allocated to candidates than the older Oxford
algorithm.

The data were taken during four separate runs. The first observations
were done in classical mode during 2 nights, 2006 March 23 and 24.
During this run the S11 field was observed. The other three occasions
were done in service mode on 2006 October 10, 2006 November 10, and
2007 March 26. During these runs both fields were targeted. We used
the 580V and the 385R volume phase holographic (VPH) gratings for the
blue and red arm, respectively. Table~\ref{tab:specobs} summarises the
observations. In total, 301 and 255 candidates were observed in the
CDFS and S11 fields, respectively.

\begin{table*}
  \begin{tabular}{lccrc}
    \hline\hline
    \centercol{observing}      & \centercol{field}      & \centercol{number of}      & \centercol{tot. exp.}  &  \\
    \centercol{dates}          & \centercol{observed}   & \centercol{configurations} & \centercol{time (sec)} & \centercol{seeing (\arcsec)}\\
    \hline
    2006/03/23 & S11  & 2 & 11,700 & 0.9--1.5\\
    2006/03/24 & S11  & 3 & 14,400 & 1.3--1.8\\
    2006/10/10 & CDFS & 1 &  9,900 & 1.8--2.2\\
    2006/11/10 & CDFS & 1 & 11,700 & 1.2--1.5\\
    2007/03/26 & S11  & 1 &  6,300 & 2.5\\
    \hline\hline
  \end{tabular}
  \caption{Details of the spectroscopic follow-up observations.}
  \label{tab:specobs}
\end{table*}

Basic spectral reductions, including bias-subtraction, flat-fielding
and wavelength calibration were done using the 2dF reduction pipeline
{\tt drcontrol}\footnotemark[\value{footnote}]. The final
one-dimensional spectrum for each object was obtained by averaging the
reduced spectra of the object in the different observations using our
own IDL scripts.

The spectra of several standard stars \citep[LTT~7379, LTT~7987 and
CD-32~9927;][]{Bessell99} were taken during the final night of the
2006 March run and were reduced in the same fashion as the science
data. System throughput as a function of wavelength was derived using
each standard star and its sensitivity curve. These curves were scaled
to a common level and averaged to give the overall sensitivity. This
was applied to all the science spectra to flux calibrate each relative
to one another. Unfortunately, absolute flux calibrations are very
difficult to do reliably with fibre-based spectrographs, due to the
changing configurations of the fibres and the effect this has on their
throughput. For this reason, we used the line fluxes measured from our
narrowband photometry rather than the fibre spectroscopy.

\subsection{Spectroscopic completeness}
\label{subsec:speccompleteness}

We used a Monte-Carlo simulation that combined the background of real
spectra of our securely confirmed \ha{} emitting galaxies with
transplanted and scaled emission lines to asses our spectroscopic
completeness as a function of line flux. We took the spectrum of each
\ha{} emitter and fitted the \ha{} and \nii{} lines together with the
continuum. Each line was fitted by a Gaussian and the galaxy
continuum (or background sky) was approximated by a first order
polynomial. The line centres were parameterised by redshift. The
widths of the \nii{} lines were set equal and the flux ratio between
the red and blue \nii{} lines was fixed to 2.96 \citep{Mendoza83}.
The remaining fit parameters were left unconstrained. The model of the
\ha{}-\nii{} complex was subtracted from our data, leaving only the
underlying noise. To the noise, we added a randomly scaled version of
our model with a random offset in wavelength. We then attempted to
re-identify any emission line. We did this multiple times for each
secure \ha{} emitting galaxy. 

\begin{figure}
  \centering
  \includeIDLfig{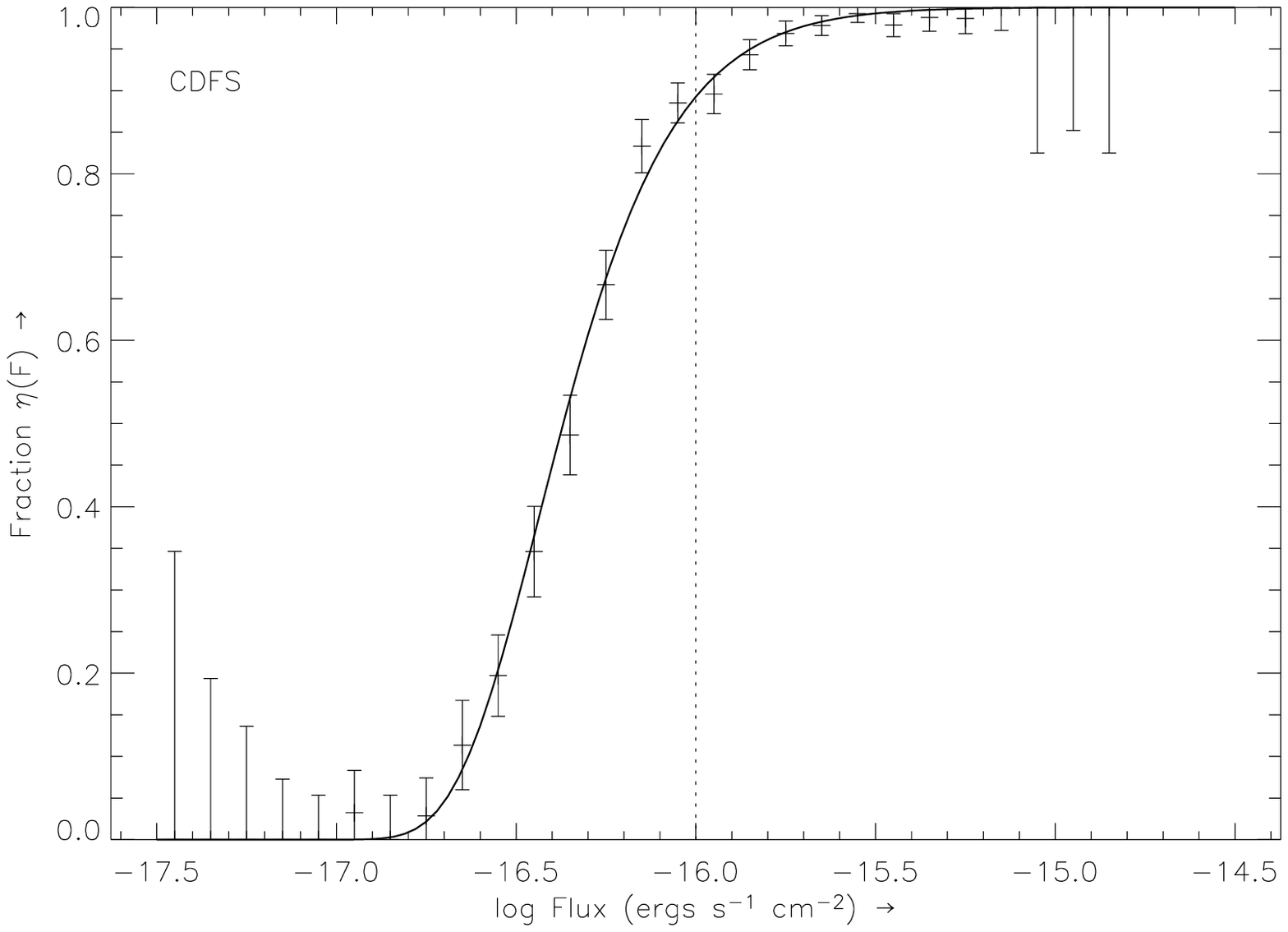}
  \includeIDLfig{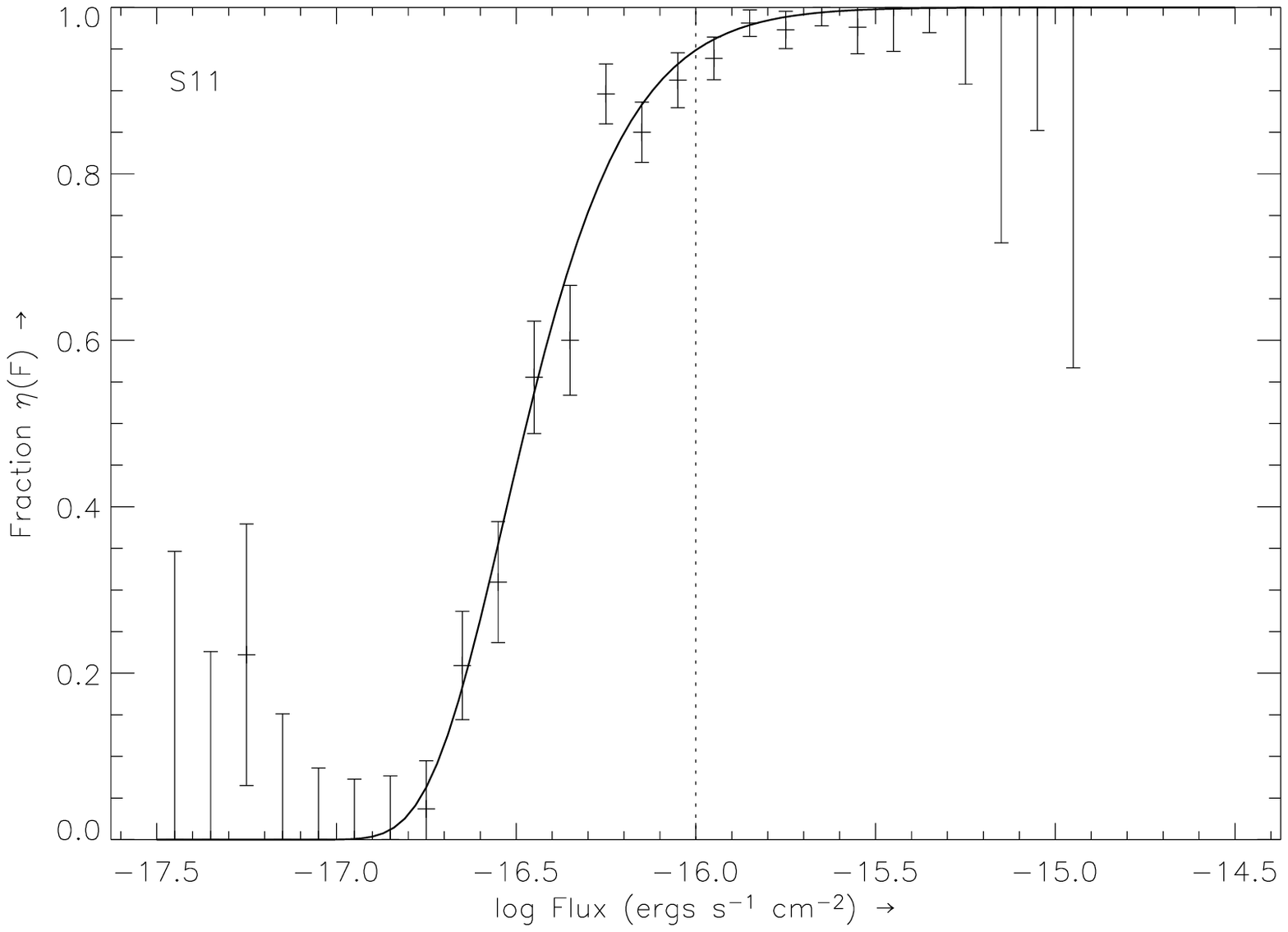}
  \caption{Spectroscopic completeness as a function of line flux for
    the CDFS ({\it top}) and S11 ({\it bottom}) fields as derived from
    a Monte-Carlo simulation of artificially generated emission lines.
    See text for details.}
  \label{fig:speccompleteness}
\end{figure}

This exercise demonstrated that it was possible to identify at least
90\,\% of the galaxies at a line flux of $\log F_\mathrm{line} =
-16.0$ ($F_\mathrm{line}$ in \lineunits{}) for all spectroscopic runs.
In Figure~\ref{fig:speccompleteness} the recovered fraction as a
function of line flux is shown for the CDFS and S11 fields. The
uncertainties indicated in Figure~\ref{fig:speccompleteness} were
derived using the following relation:
\begin{equation}
  \sigma_\mathrm{frac}  = \frac{
    \sqrt{ N_\mathrm{tot} ( N_\mathrm{det} + 2 )(N_\mathrm{tot} -
      N_\mathrm{det} + 1)}}{N_\mathrm{tot} ( N_\mathrm{tot} + 3)}~,
  \label{eq:countUncert}
\end{equation}
where $\sigma_\mathrm{frac}$ is the calculated uncertainty,
$N_\mathrm{tot}$ the total number of objects in that bin and
$N_\mathrm{det}$ is the number of objects which have a detection of
the emission line \citep[after Eq.~4 from][]{Jones06}. The
spectroscopic completion rate as indicated in
Figure~\ref{fig:speccompleteness} is well fit by a function of the
form
\begin{equation}
  \eta (F) = 
  \left \{ \begin{array}{ll}
      \exp [-\gamma (F - F_c)^{20}] & F < F_c\\
      1 & F \ge F_c
      \end{array}
      \right.~,
  \label{eq:speccompl}
\end{equation}
where $\gamma$ represents the speed at which the function drops off
and $F_c$ is the flux at which the function reaches $1.0$.

\subsection{\ha{} emission line fraction}
\label{subsec:hafrac}
\begin{figure*}
  \centering
  \includegraphics[width=\textwidth, trim=20pt 98pt 5pt 100pt, clip]{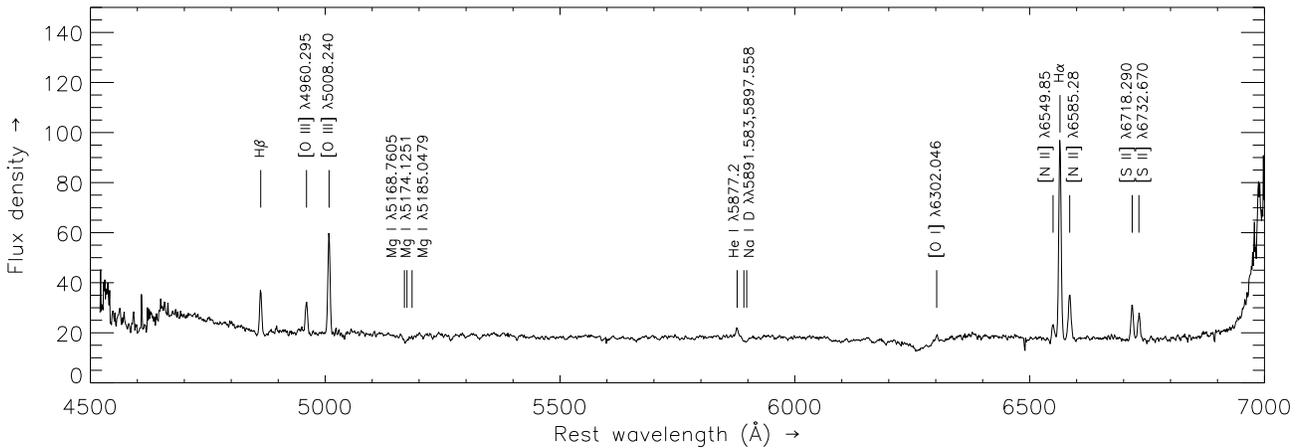}
  \caption{Mean spectrum of emission-line galaxies from the CDFS field. Spectra from
    114 galaxies between $z=0.19$ and 0.27 were de-redshifted before
    stacking. The most prominent features have been labelled. The
    apparent absorption feature just bluewards of the \Oi{} line is
    the remnant of the telluric A-band of the individual spectra being
    de-redshifted and stacked. This spectrum was used to fit the
    emission lines to derive the mean extinction as described in
    Section~\ref{subsec:extinction}. Only red arm data from AAOmega
    (observed wavelength $\sim5700-8700$\,\AA{}) are shown.}
  \label{fig:stackSpectrum}
\end{figure*}

In almost all cases the spectra of confirmed emission-line galaxies
should show additional emission lines elsewhere except cases of \lya{}
at $z\sim5.7$ (which are filtered out through their absence of $B$ and
$R$ flux) or \oii{} at $z\sim1.2$. This is demonstrated by
Figure~\ref{fig:stackSpectrum}, where we show the stacked spectrum of
all our confirmed \ha{} and \sii{} galaxies in the CDFS. \ha{} is
usually accompanied by the \Nii{} and \Sii{} doublets, whereas \hb{}
and the \Oiii{} doublet are almost always seen together. Our spectral
resolution ($R\sim1500$ at 8150\,\AA{}) is not enough to fully resolve
the \oii{} doublet, but high enough to show it as broader than a
single emission line. The signal-to-noise ratio of the spectra is not
always high enough to clearly determine if a line is broad (in this
sense) or not. Alternatively, these galaxies could be \ha{} emitting
galaxies with all other emission lines too faint to be detected.

There are a few galaxies which show only one emission line. Although
we expect many of them to be \oii{} emitters at $z\sim1.2$, we cannot
rule out the possibility of single-line \ha{} galaxies at $z\sim0.24$
in without the use of additional information. In
Figure~\ref{fig:BRcol_histo_elg} we show the observed \BRcol{} colour
distribution of galaxies in the CDFS where the emission line in the
narrowband filters has been confirmed as \ha{} or \sii{} through the
presence of additional lines. We also indicate the colour distribution
of galaxies for which we have only one emission line feature. Some of
the single-line detections are bluer than the combined \ha{}/\sii{}
distribution. We therefore identify all single-line galaxies with
\BRcol{}\,$\le$\,0.5 to be \oii{} emitters at $z\sim1.2$ and those
with \BRcol{}\,$>$\,0.5 to be \ha{} emitters at $z\sim0.24$.

Of the candidates for which we have spectroscopically confirmed an
emission line (189 and 117 out of the total 301 and 255 observed in
the CDFS and S11 fields, respectively), just under a half are \ha{} at
$z\sim0.24$, a quarter are \sii{} at $z\sim0.21$, roughly a sixth are
\hb{} or \oiii{} at $z\sim0.6-0.7$ and the remainder are \oii{} at
$z\sim1.2$.

\begin{figure}
  \centering
  \includeIDLfig{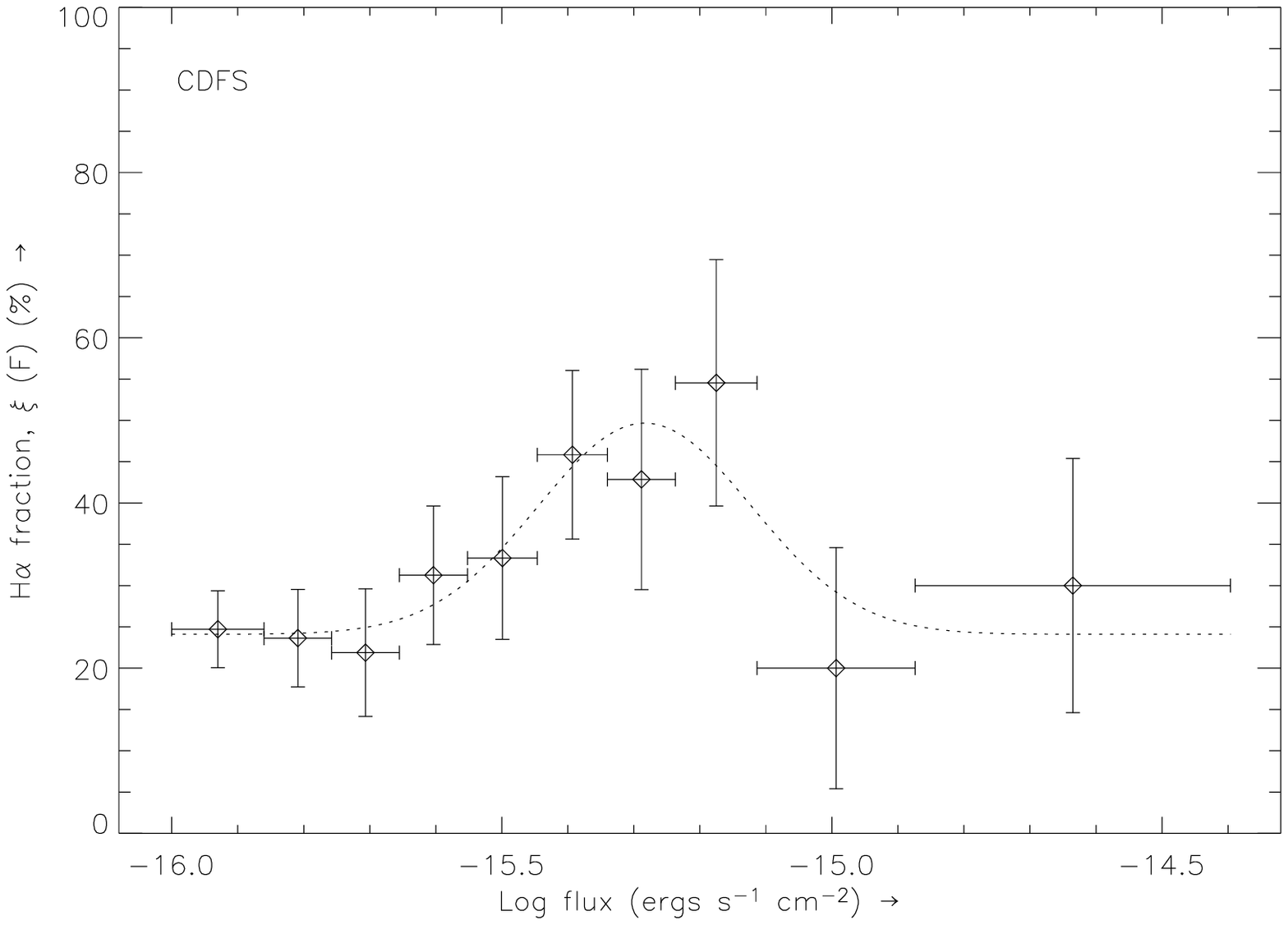}
  \includeIDLfig{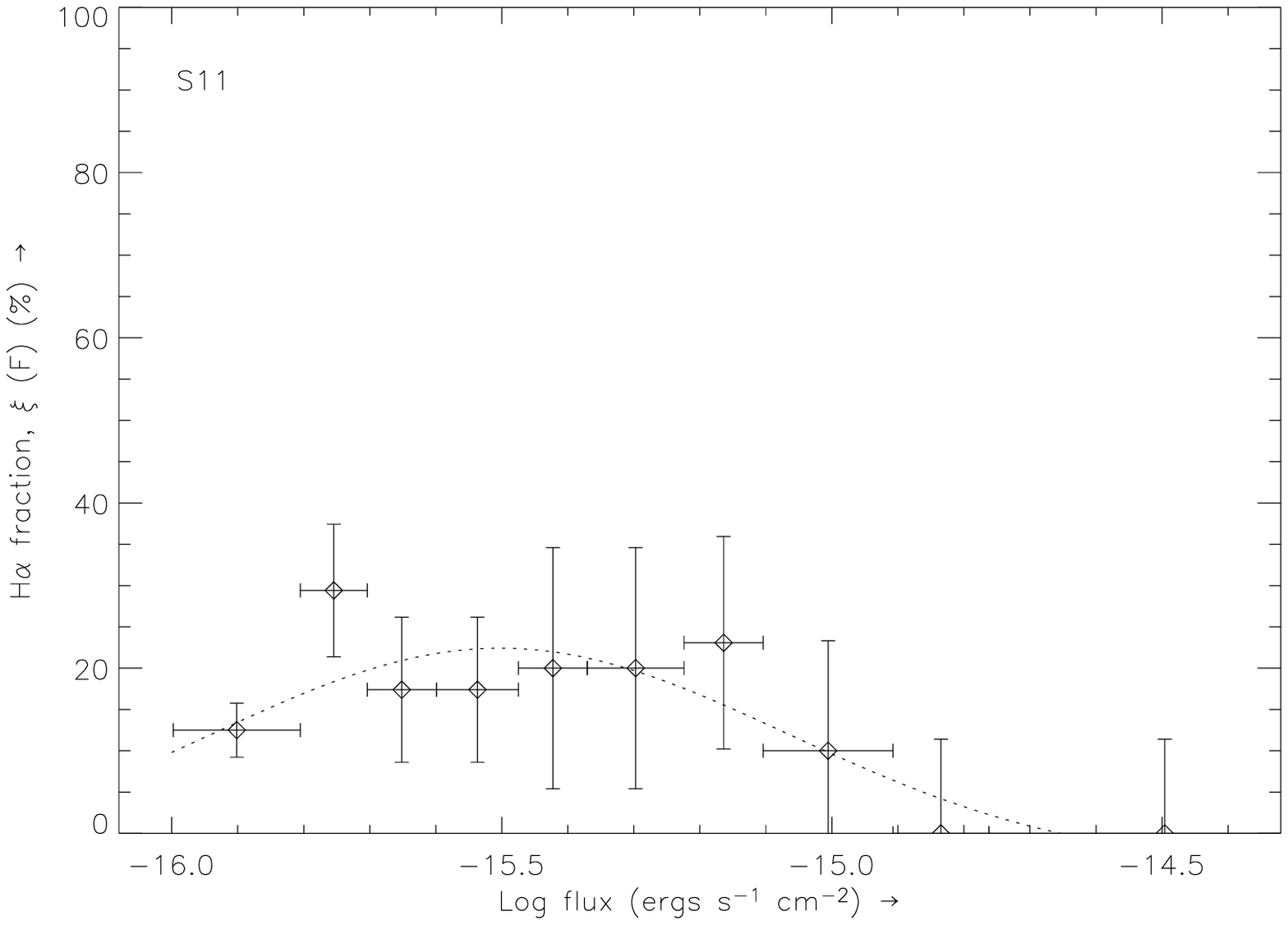}
  \caption{The \ha{} fraction of our candidates for the CDFS ({\it
      top}) and S11 ({\it bottom}) fields. The data have been binned to
    have a minimum of 10 galaxies per bin and a minimum width of
    0.1\,dex. The dotted line is the four-parameter fit to the data
    points. The horizontal error-bars represent the width of the bins
    and the vertical bars the uncertainty in the fraction calculated
    using Eq.~\eqref{eq:countUncert}.}
  \label{fig:haFrac}
\end{figure}

Figure~\ref{fig:haFrac} shows the fraction of confirmed \ha{} emitters
in our full spectroscopic sample as a function of narrowband flux. It
peaks around $\log F_\mathrm{line} \sim -15.3$ (with $F_\mathrm{line}$
in \lineunits{}), below which increasing numbers of \oii{} galaxies at
$z\sim1.2$ begin to dominate the counts. Each point in
Figure~\ref{fig:haFrac} has a minimum of 10 galaxies per bin and a
minimum binwidth of 0.1\,dex. The uncertainties in the \ha{} fraction
per bin have been calculated using Eq.~\ref{eq:countUncert}, where
$N_\mathrm{det}$ now represents the number of galaxies with confirmed
\ha{}. We fit a Gaussian of the form
\begin{equation}
\xi (F) = a \times \mathrm{exp} \left (- \frac{(F - F_c)^2}{2 \sigma^2} \right ) + b~,
\end{equation}
where $F_c$ is the flux central to the peak, $\sigma$ and $a$ are its
width and height, and $b$ is a zero-point offset. The resulting fits
are shown in Figure~\ref{fig:haFrac}.

We decided to fit both the CDFS and S11 fields individually, given the
likely differences between the field samples due to cosmic variance.
Given the relatively narrow range of volume probed through each
emission line, we expect over- and underdensities at the different
redshift intervals to change the relative numbers of galaxies as a
function of flux \citep{Jones01,Pascual01}.

\subsection{Extinction corrections}
\label{subsec:extinction}
Star forming regions are some of the dustiest galaxy environments,
making correction for internal obscuration necessary. Many emission
line surveys apply a general extinction correction of
$A_\mathrm{\ha{}}$\,$\sim$\,1 \citep[e.g.][]{Tresse98,Fujita03}.
However, it has been shown that there are large variations in
extinction between galaxies \citep[e.g.][]{Jansen01}. Furthermore,
\citet{Massarotti01} state that applying an average extinction
correction always underestimates the true extinction correction. Since
our spectra cover a large wavelength range (3800--5700\,\AA{} in the
blue and 5700--8700\,\AA{} in the red) we are able to observe \ha{}
and \hb{} simultaneously. We therefore calculate the extinction
individually for each galaxy through \ha{} and \hb{} when both lines
are detectable. The signal-to-noise ratio is not always high enough to
show \hb{} clearly in emission. Therefore, we grouped available
spectra according to the $B$-magnitude of the source, obtained an
average spectrum, and measured the Balmer decrement value from these.

\begin{figure}
  \centering
  \includeIDLfig{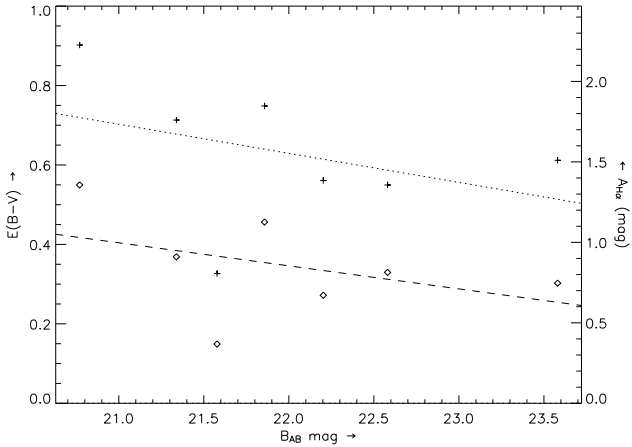}
  \caption{Colour excess E(B-V) and $A_\mathrm{\ha{}}$ (assuming
    $k(\mathrm{\ha{}})$ = 2.47) as a function of $B$-magnitude as
    determined by measuring the Balmer decrement of the averaged
    spectra in the CDFS. The crosses indicate the colour excess
    without using any correction for stellar absorption in \ha{} and
    \hb{}. The diamonds indicate the colour excess using the same
    stellar absorption correction as \citet{Hopkins03} with
    EW(\ha{})\,=\,1.3 and EW(\hb{})\,=\,1.6. The dotted and dashed
    line are the linear fits to the respective points.}
  \label{fig:EBminV}
\end{figure}

The colour excess $E(B-V)$ can be calculated using
\begin{equation}
  E(B-V)= \frac{2.5 \log R_{\alpha\beta}}{k(\mathrm{\hb{}}) - k(\mathrm{\ha{}})}~,
\end{equation}
where $R_{\alpha\beta}$ is the ratio of the observed value of the
Balmer decrement to its theoretical value, and $k(\mathrm{\hb{}}) -
k(\mathrm{\ha{}})$ is the differential extinction between the
wavelengths of \hb{} and \ha{}. The theoretical value for the Balmer
decrement is 2.87 \citep[for $T\,=\,\pow{4}$\,K and case B
recombination; Table~2 of][which uses a \citealp{Cardelli89}
extinction law]{Calzetti01} and the value for the differential
extinction is 1.163. This assumes $k(V)$\,=\,3.1 and
$k(\ha{})$\,=\,2.468. We adopt these values throughout the rest of
this paper.

In Figure~\ref{fig:EBminV} we plot the resulting values for $E(B-V)$
as a function of the $B$-magnitude for two cases: without and with
correction for absorption due to the underlying stellar population.
The AAOmega spectra have a resolution of $\sim$\,5.3\,\AA{} throughout
the red arm meaning that we are unable to resolve the \hb{} absorption
line directly. If we assume no stellar absorption, the colour excess
has values up to $E(B-V)\sim1$, corresponding to
$A_\mathrm{\ha{}}$\,=\,2.5\,mag (or $A_\mathrm{V}$\,=\,3.1\,mag) using
$A_\lambda = k(\lambda) \times E(B-V)$. This is far higher than the
average extinction of $A_\mathrm{\ha{}}$\,$\sim$\,1 as assumed
elsewhere \citep[e.g.][]{Tresse98,Fujita03}. If we instead adopt the
median equivalent widths for stellar absorption in \ha{} and \hb{} as
measured by \citet{Hopkins03}, 1.3 and 1.6\,\AA{} respectively, then
the average extinction as shown in Figure~\ref{fig:EBminV} is roughly
$A_\mathrm{\ha{}}$\,$\sim$\,0.85. We note that there is a trend of a
decreasing extinction with increasing apparent magnitude (see
Figure~\ref{fig:EBminV}). Observe that our sample has a restricted
range in redshift, making apparent magnitude $B$ a proxy for absolute
magnitude $M_B$. Similar trends of change in $E(B-V)$ have been found
by \citet{Jansen01}. We attribute this trend to the fact that either
fainter (and therefore smaller) galaxies potentially contain less
dust, or the \hb{} flux might be overestimated in the mean spectrum of
the faintest galaxies as a result of a low signal-to-noise ratio of
the \hb{} line. We derive an extinction of $A_\mathrm{\ha{}} =
0.96$\footnote{Alternatively, the extinction law of \citet{Calzetti00}
  gives $A_\mathrm{\ha{}} = 1.18$, where $k(V)$\,=\,4.05,
  $k(\ha{})$\,=\,3.325 and $k(\mathrm{\hb{}}) - k(\mathrm{\ha{}}) =
  1.163$. This correction results in \ha{} fluxes roughly 20\,\%
  higher than the values used in the text.}  from the Balmer ratio in
the mean spectrum of all emission-line galaxies as shown in
Figure~\ref{fig:stackSpectrum}. Since the trend might be due to a low
signal-to-noise ratio of the \hb{} line, we use a constant value
throughout to correct for extinction.

\section{Luminosity function and star formation density}
\label{sec:lumfieSFD}
\subsection{Derivation and fit}
\label{subsec:schechterfitting}

\begin{figure*}
  \centering
  \includegraphics[width=\columnwidth, trim=15pt 8pt 5pt 18pt, clip]{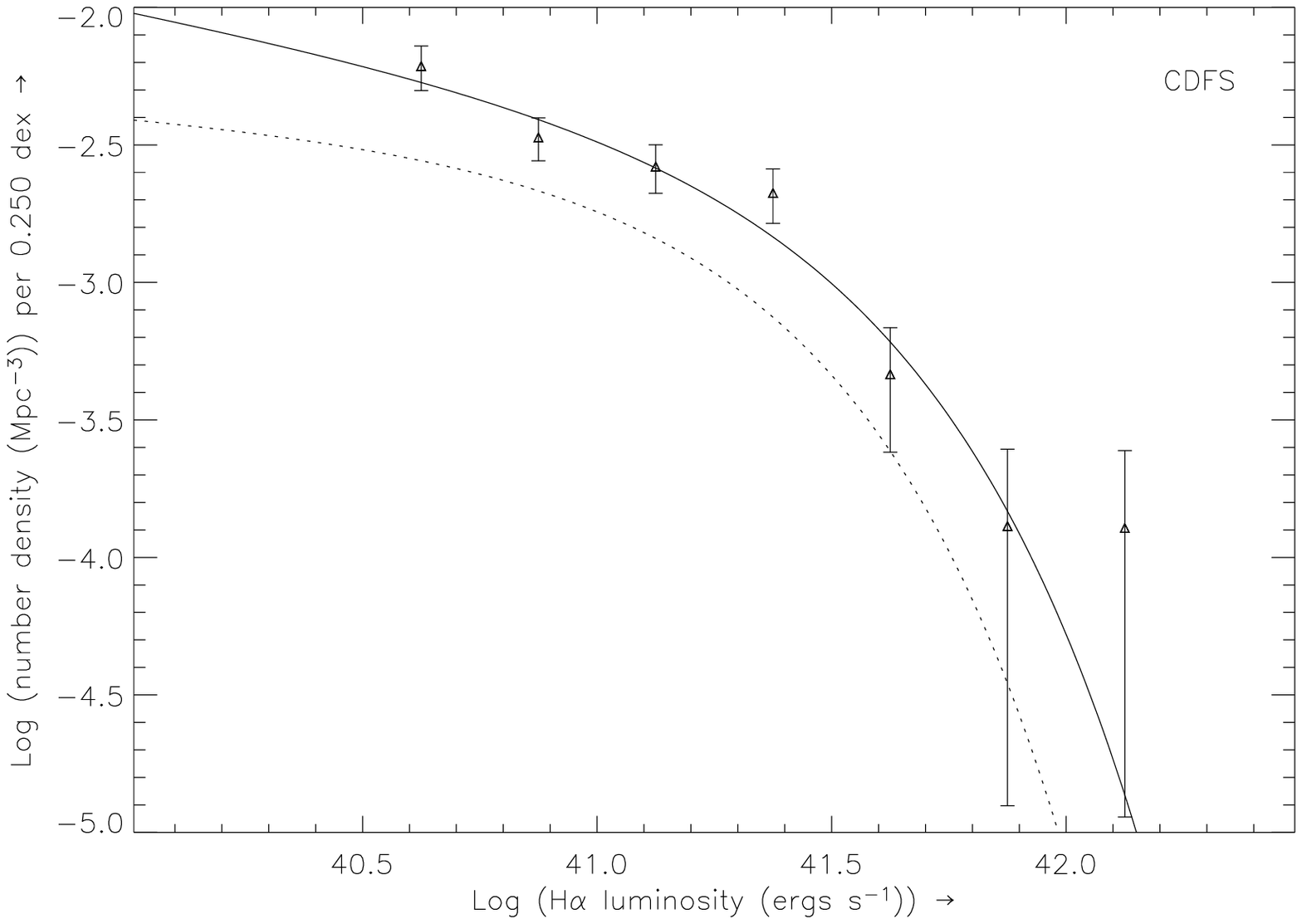}
  \includegraphics[width=\columnwidth, trim=15pt 8pt 5pt 18pt, clip]{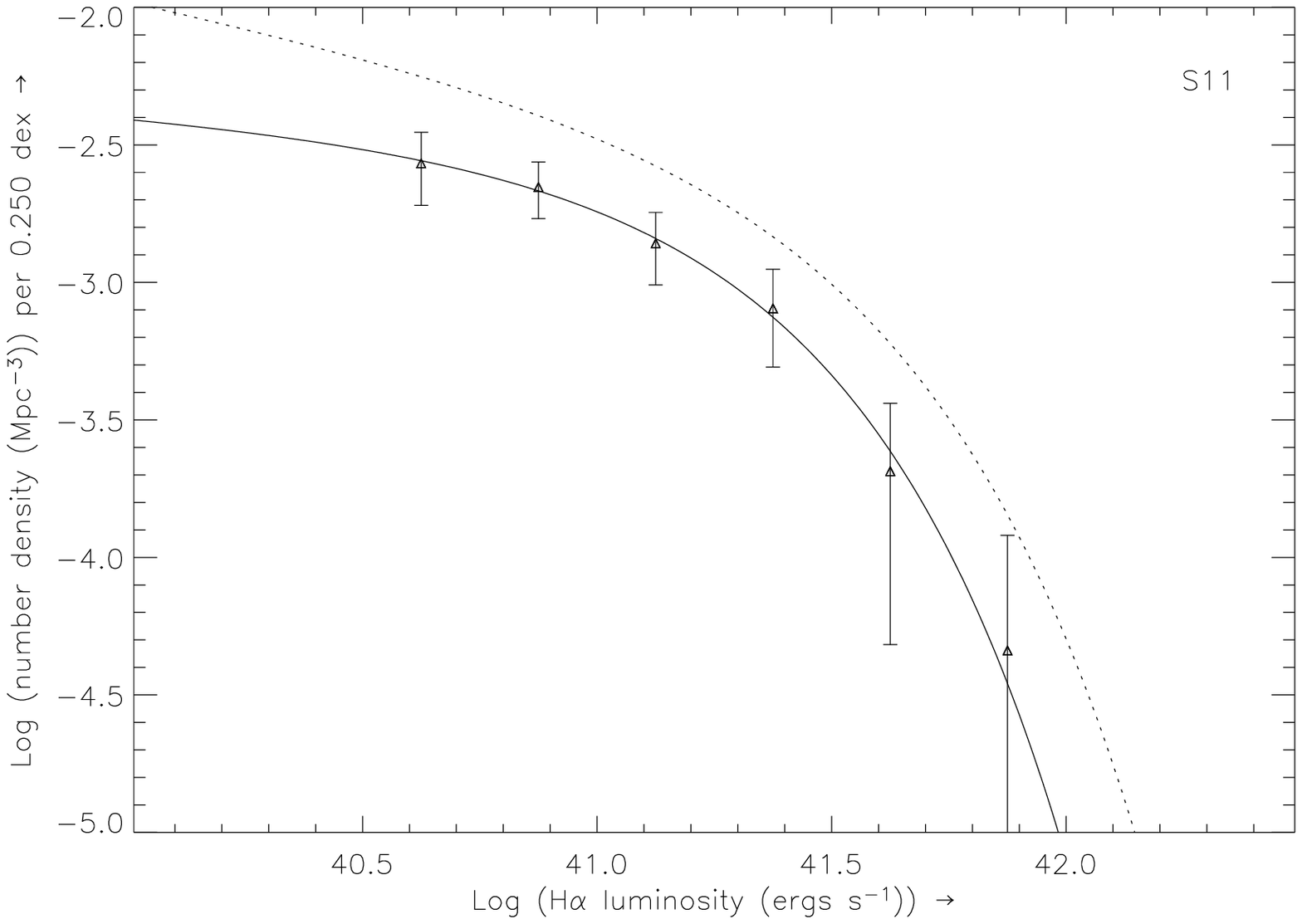}
  \includegraphics[width=\columnwidth, trim=20pt 8pt 5pt 2pt, clip]{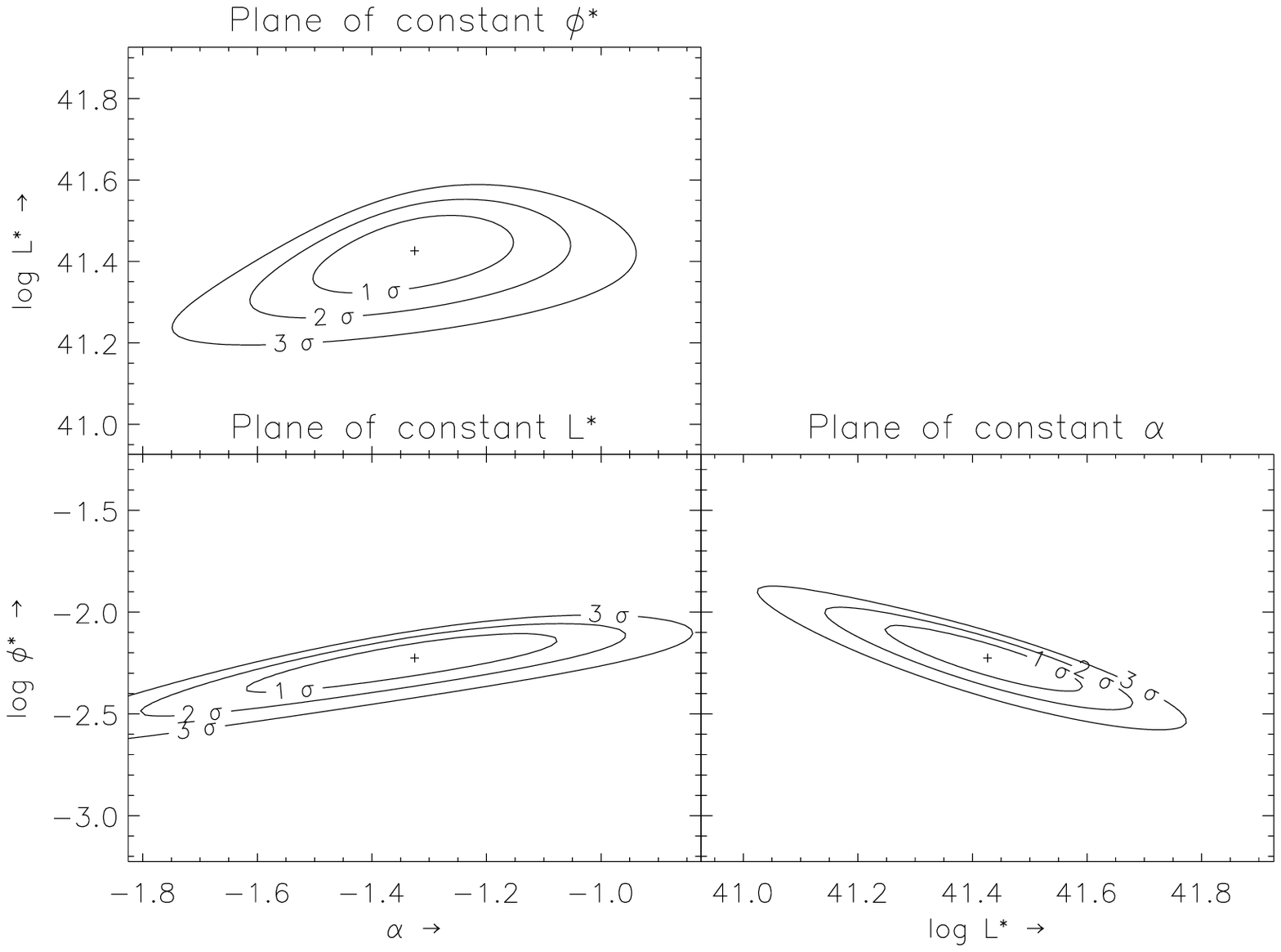}
  \includegraphics[width=\columnwidth, trim=20pt 8pt 5pt 2pt, clip]{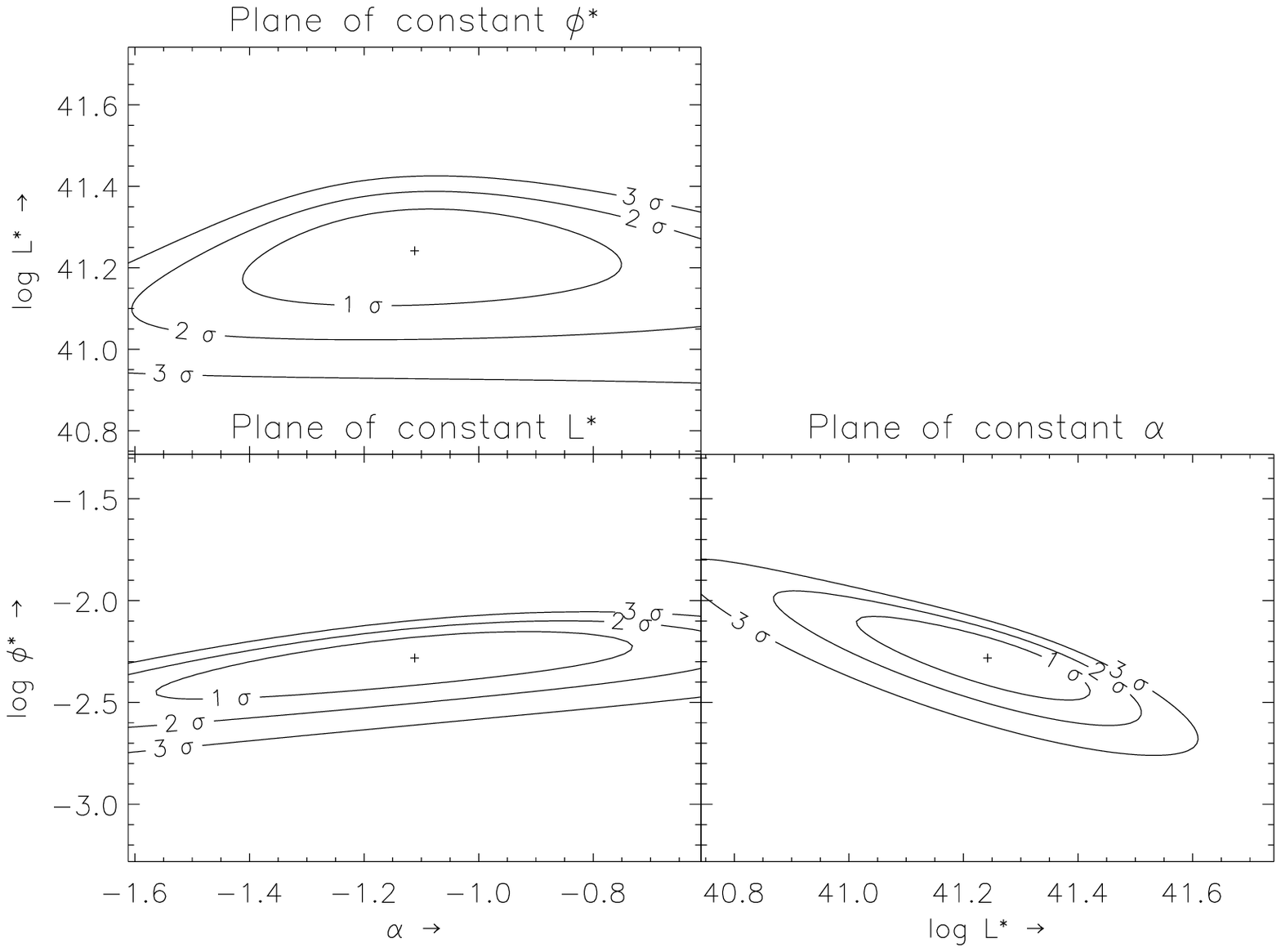}
  \caption{{\it Top}: luminosity function for \ha{} galaxies at
    $z\sim0.24$ for the CDFS ({\it left}) and S11 ({\it right})
    fields. The solid line in each of these panels is the fit to the
    data points, while the dotted line indicates the fit of other
    field for reference. {\it Bottom}: Confidence levels for the
    parameters $\alpha$, $L^*$ and $\phi^*$ of the CDFS ({\it left})
    and S11 ({\it right}) fields. Contours are drawn for each plane in
    which one of the parameters is held constant. The 1, 2 and
    3\,$\sigma$ contours indicated correspond to 68.3\%, 95.4\% and
    99.7\% confidence limits, respectively.}
  \label{fig:halumfie}
\end{figure*}

With the final emission line catalogue in hand, and the various
selection and completeness effects accounted for, our approach to
calculating the \ha{} luminosity function is as follows. We take our
measured distribution of line emitters (all emission lines from all
redshifts) from the narrowband candidate sample and apply the
spectroscopically measured fraction of \ha{} emitters as a function of
flux (Section~\ref{subsec:hafrac}). We correct for incompleteness in
both the spectroscopic identifications
(Section~\ref{subsec:speccompleteness}) as well as the original
narrowband imaging. The corrections for the latter are less than
0.1\,\% (Section~\ref{sec:photcompcorr}). Finally, we correct our
line fluxes for the effects of extinction
(Section~\ref{subsec:extinction}).

Figure~\ref{fig:halumfie} shows separate luminosity functions for both
the CDFS and S11 fields. We fit a Schechter function
\citep{Schechter76} to the data points using a minimised $\chi^2$ fit.
The Schechter function is given by
\begin{equation}
  \phi (L) dL = \phi^* \left ( \frac{L}{L^*} \right ) ^ {-\alpha} \exp
  \left ( - \frac{L}{L^*} \right ) d \left ( \frac{L}{L^*} \right )~,
  \label{eq:schechter}
\end{equation}
where $\phi^*$ represents the normalisation constant of the galaxy
density, $\alpha$ the faint end slope, and $L^*$ the characteristic
luminosity where the Schechter function rapidly declines at bright
luminosities. We used a Levenberg-Marquardt method for finding the
minimum $\chi^2$ fit to the binned data-points, courtesy of the
IDL routine {\tt mpfitfun} from the Markwardt\footnote{Maintained by
  C.~Markwardt at http://cow.physics.wisc.edu/$\sim$craigm/\linebreak
  idl/idl.html.} library. Since the three parameters $\alpha$, $L^*$
and $\phi^*$ are highly correlated, we used the correlation matrix and
the partial derivatives of the Schechter function to calculate the
formal uncertainty in the integrated luminosity density $\mathcal{L}$,
\begin{equation}
\sigma^2_\mathcal{L} = \sum \limits ^3 _{i,j=1} \left [\frac{\partial
    \mathcal{L}}{\partial x_i} \frac{\partial
    \mathcal{L}}{\partial x_j} \right ]_{x=\mu} V_{ij} ~.
\end{equation}
Here, $x_1$, $x_2$ and $x_3$ correspond to the Schechter parameters
$\alpha$, $\log L^*$ and $\log \phi^*$ \citep{Cowan98}. $V_{ij}$ is
the covariance matrix, which relates to the correlation matrix
$\rho_{ij}$ as $V_{ij} = \rho_{ij} \sigma_i \sigma_j$. $\sigma_i$ is
the formal uncertainty in the $i^\mathrm{th}$ parameter. We list the
resulting values of the parameters and the formal uncertainties,
together with the correlation matrices in
Table~\ref{tab:schechterfit}.

\begin{table*}
  \centering
  \begin{tabular}{ccccccc}
    \hline\hline
    \multicolumn{3}{c}{CDFS} & & \multicolumn{3}{c}{S11}\\
    \cline{1-3}\cline{5-7}
    $\alpha$ & $\log L^*$ & $\log \phi^*$ & &
    $\alpha$ & $\log L^*$ & $\log \phi^*$\\
    $-$1.33 $\pm$ 0.34 & 41.43 $\pm$ 0.22 & $-$2.23 $\pm$ 0.32& &
    $-$1.11 $\pm$ 0.51 & 41.24 $\pm$ 0.25 & $-$2.28 $\pm$ 0.33\\
    \cline{1-3}\cline{5-7}
    \multicolumn{3}{c}{
      $ \left ( \begin{array}{rrr}
           1.00000 & -0.91020 &  0.96458 \\
          -0.91020 &  1.00000 & -0.97268 \\
           0.96458 & -0.97268 &  1.00000 
         \end{array} \right ) $
    } & &
    \multicolumn{3}{c}{
      $ \left ( \begin{array}{rrr}
           1.00000 & -0.90948 &  0.95099 \\
          -0.90948 &  1.00000 & -0.96826 \\
           0.95099 & -0.96826 &  1.00000
        \end{array} \right ) $
    }\\
    \hline\hline
  \end{tabular}
  \caption{Schechter parameters for the \ha{} luminosity functions for
    each field determined using a Levenberg-Marquardt $\chi^2$
    minimisation. The correlation matrices $\rho_{ij}$ for each are
    shown below.}
  \label{tab:schechterfit}
\end{table*}

The luminosity density over luminosities $L \ge L_\mathrm{lim}$ can be
calculated by integrating Eq.~\ref{eq:schechter}, yielding
\begin{equation}
  \mathcal{L} = \phi^* L^* \Gamma (\alpha + 2,
  \frac{L_\mathrm{lim}}{L^*})~.
  \label{eq:lumintegration}
\end{equation}
In the case where limiting luminosity $L_\mathrm{lim} = 0$, the
luminosity density reduces to $\mathcal{L} = \phi^* L^* \Gamma (\alpha
+ 2)$. Using the Schechter parameters and uncertainties given in
Table~\ref{tab:schechterfit} with $\log L_\mathrm{lim}=40.6$
($L_\mathrm{lim}$ in \ergs{}, corresponding to our survey flux limit)
gives $\log \mathcal{L} = 39.17 ^{+0.08} _{-0.10}$ and $38.86 ^{+0.11}
_{- 0.14}$ in \ergs{} for the CDFS and S11 fields, respectively. The
uncertainties are calculated using the correlation matrices in
Table~\ref{tab:schechterfit}. If we instead use the \ha{} luminosities
of the galaxies directly and sum over all, we obtain $39.22 ^{+0.02}
_{-0.02}$ and $38.86 ^{+0.03} _{-0.03}$ for CDFS and S11,
respectively. The uncertainties in this case are the square-root of
the sum in quadrature of individual galaxy luminosity uncertainties
and does not take into account \ha{} emission line fraction
uncertainties and, as such, are lower limits.

\subsection{Comparison to previous surveys}
\label{subsec:comp}
\begin{figure}
  \centering
  \includegraphics[width=\columnwidth, trim=15pt 8pt 5pt 18pt, clip]{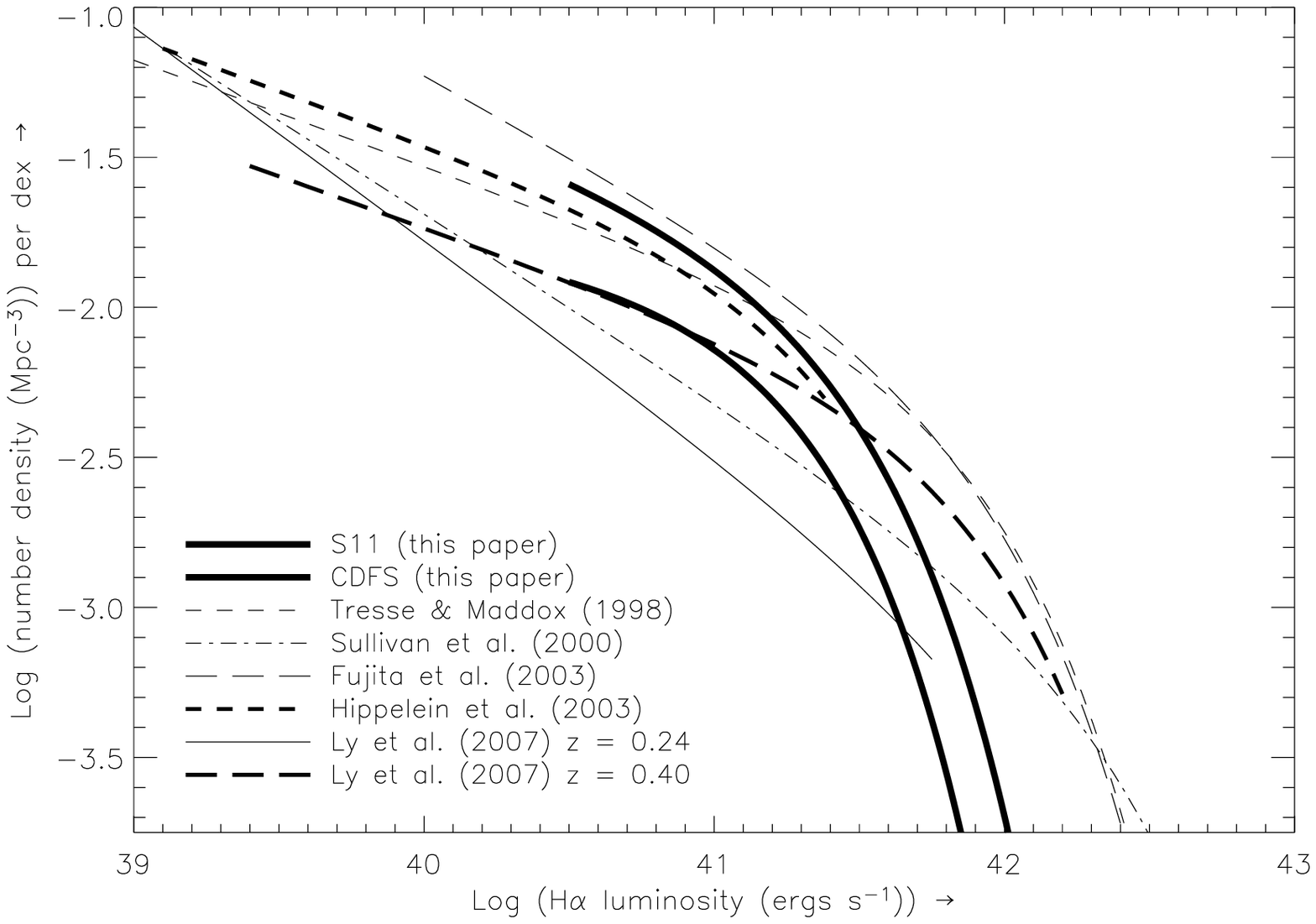}
  \caption{Comparison of the Schechter functions derived for our two
    fields (thick solid lines) and those of other surveys. The
    Schechter function of each survey has been drawn over the
    luminosity range where data was available. The other surveys are
    \citet[dash]{Tresse98}, \citet[dash dot]{Sullivan00}, \citet[long
    dash]{Fujita03}, \citet[thick short dash]{Hippelein03}, and
    \citet[ $z=0.24$ solid; $z=0.40$ thick long dash]{Ly07}. The
    individual Schechter parameters are given in
    Table~\ref{tab:surveypars}.}
  \label{fig:schechtercomparison}
\end{figure}

In Figure~\ref{fig:schechtercomparison} we compare our Schechter fits
to the results of other surveys using \ha{} as a measure for star
formation. The survey parameters are summarised in
Table~\ref{tab:surveypars}. We restricted the comparison to \ha{}
surveys with $z\lesssim0.40$ in order to limit the systematic
uncertainties which play into the comparison when different star
formation indicators are involved. It can be seen that there is a
large range in each of the Schechter parameters between surveys.
$\alpha$ ranges from $\sim -1.1$ to $-1.6$, $\log L^*$ from $\sim41.3$
to $42.2$ and $\log \phi^*$ from $-3.7$ to $-2.2$. Some of these
surveys cover different redshifts to those in our survey. The wide
span of the parameters could be attributed by evolution of the
luminosity function, as has been suggested by \citet{Hopkins04} and
\citet{Ly07}, who compare surveys over a wider redshift range using
different indicators. However, a number of systematic uncertainties
exist between surveys that could also attribute to the scatter between
the luminosity functions. We now explore each in turn.

\begin{table*}
  \centering
  \begin{tabular}{lcccc}
    \hline\hline
    \centercol{Reference/field} & \centercol{Redshift} & \centercol{$\alpha$} & \centercol{$\log L^*$} & \centercol{$\log \phi^*$}\\
    \hline
    \citet{Gallego95}   & 0.022 $\pm$ 0.022 & $-$1.30 $\pm$ 0.20 & 41.87 $\pm$ 0.08 & $-$2.79 $\pm$ 0.20\\
    \citet{Tresse98}    & 0.200 $\pm$ 0.100 & $-$1.35 $\pm$ 0.06 & 41.92 $\pm$ 0.13 & $-$2.56 $\pm$ 0.09\\
    \citet{Sullivan00}  & 0.150 $\pm$ 0.150 & $-$1.62 $\pm$ 0.10 & 42.42 $\pm$ 0.14 & $-$3.55 $\pm$ 0.20\\
    \citet{Fujita03}    & 0.242 $\pm$ 0.009 & $-$1.53 $\pm$ 0.15 & 41.95 $\pm$ 0.25 & $-$2.62 $\pm$ 0.34\\
    \citet{Hippelein03} & 0.245 $\pm$ 0.022 & $-$1.35            & 41.45            & $-$2.32           \\
    \citet{Perez03}     & 0.025 $\pm$ 0.025 & $-$1.20 $\pm$ 0.20 & 42.43 $\pm$ 0.17 & $-$3.00 $\pm$ 0.20\\
    \citet{Ly07}        & 0.0735 $\pm$ 0.0075, 0.0855 $\pm$ 0.0055 & $-$1.59 $\pm$ 0.02 & 42.05 $\pm$ 0.07 & $-$3.14 $\pm$ 0.09\\
    \citet{Ly07}        & 0.242 $\pm$ 0.009 & $-$1.71 $\pm$ 0.08 & 42.20 $\pm$ 1.24 & $-$3.70 $\pm$ 1.06\\
    \citet{Ly07}        & 0.401 $\pm$ 0.010 & $-$1.34 $\pm$ 0.06 & 41.93 $\pm$ 0.19 & $-$2.75 $\pm$ 0.16\\
    \\
    This paper, CDFS    & 0.245 $\pm$ 0.016 & $-$1.33 $\pm$ 0.34 & 41.43 $\pm$ 0.22 & $-$2.23 $\pm$ 0.32\\
    This paper, S11     & 0.245 $\pm$ 0.016 & $-$1.11 $\pm$ 0.51 & 41.24 $\pm$ 0.25 & $-$2.28 $\pm$ 0.33\\
    \hline\hline
  \end{tabular}
  \caption{Values for the parameters of the Schechter functions shown
    in Figure~\ref{fig:schechtercomparison}. $L^*$ is in \ergs{} and
    $\phi^*$ in \perMpc{}. After Table~5 in
    \citet{Ly07}.}
  \label{tab:surveypars}
\end{table*}

The details of galaxy selection inevitably vary from survey to survey.
For example, \citet{Tresse98} have selected their galaxies from an
$I$-band selected sample, while \citet{Sullivan00} used UV imaging to
select theirs. It is well known that galaxy selection based on
different passbands results in a different faint end slope of the
galaxy luminosity distribution \citep[][]{Madgwick02,Jones04}.
Passbands that favour bluer and/or star forming galaxies generally
yield higher faint end counts and thus steeper slopes. This
undoubtedly has a similar influence on the faint end slope of the
\ha{} luminosity function.

It is also important to note that any survey using an equivalent width
selection \citetext{or equivalently, a narrowband$-$broadband colour,
  e.g. \citealp{Fujita03} and \citealp{Ly07}}, unlike our survey which
rather applies a flux limit, will tend to be biased against galaxies
with low equivalent widths. This will affect mostly the selection of
galaxies with a high star formation rate per unit continuum, such as
early type spirals \citep{Kennicutt92} and galaxies with low \ha{}
flux in general. The \ha{} luminosity function, of course, only
characterises the line flux on its own.

In Section~\ref{subsec:extinction} we discussed the amount of
extinction correction for our survey and concluded that it agrees with
values found by other surveys. However, there is still a large spread
in the extinction values. A range of $A_\mathrm{\ha{}} = 0.5-1.8$ is
typical of those found \citep[][and references
therein]{Ly07,Kennicutt98}, which translates directly into an
uncertainty of 0.3 in $\log L^*$. The exception is when all galaxies
have individually been corrected for extinction, which imposes large
observational overheads. None of the surveys indicated in
Figure~\ref{fig:schechtercomparison} have been able to do so.

Some surveys have only a limited spectroscopic follow up on their
candidates, or none at all \citep{Fujita03,Ly07}. Both of these
surveys use additional colour criteria to distinguish between
\ha{} and other line emitting galaxies at other redshifts.
\citet{Ly07} estimate that there is about 50\,\% contamination of
\oiii{} galaxies into the \ha{} sample of \citet{Fujita03} based on
empirical colour selection using spectra from the Hawaii Hubble Deep
Field-North. Spectroscopy on several sources in \citet{Ly07} shows
that slight contamination of higher redshift emission line galaxies
occurs in their \ha{} sample. In our own sample, as we noted in
Section~\ref{subsec:spectroscopy}, there is a large sample of galaxies
that has been selected on their \sii{} lines, which would have
otherwise been mistaken for low redshift \ha{} had we relied on colour
selection on its own. Furthermore, the fraction of contamination by
other emission line galaxies varies significantly with observed line
flux \citep{Jones01,Pascual01}. Hence, spectroscopic observations of
all or a large representative sample of the candidates is vital in
understanding the amount of contamination by galaxies at different
redshifts.

Spectroscopic observations also allow flux corrections for the \Nii{}
lines, which straddle \ha{} with an observed separation of
$\sim44$\,\AA{} at $z\sim0.24$. A more detailed analysis of \nii{} is
described in Section~\ref{subsec:nii}. Fluxes quoted in this paper do
not include this correction, unless otherwise stated.

Cosmic variance has widely been cited as a major contributor to the
differences between various surveys \citep[e.g.][]{Ly07}. We are
well-placed to test the impact of this given that we have observed two
distinct fields that have been subjected to identical selection and
analysis. We have estimated the contribution of cosmic variance to the
mean object densities given by the luminosity functions in
Figure~\ref{fig:schechtercomparison}. Following the prescription of
\citet{Somerville04} we determined the relative cosmic variance
$\sigma_v^2$ for several \ha{} surveys. The estimate of $\sigma_v$ is
an upper limit as our survey has the shape of an elongated prism,
while the derivation is for a spherical volume \citep{Somerville04}.
The cosmic variance is calculated by $\sigma_v = b
\sigma_\mathrm{DM}$, where $b$ is the bias parameter (defined as the
ratio of the root variance of the halos and the dark matter) and
$\sigma_\mathrm{DM} ^2$ the variance of the dark matter. Using a
number density of 0.05\,\perMpc{} \citep{Ly07} yields a bias of
$b\sim0.7$ for all surveys at $z\lesssim0.40$. The corresponding
variance over our survey volumes (of $9.4\times\pow{3}$ and
$8.3\times\pow{3}$\,\Mpc{}) is $\sigma_\mathrm{DM} \sim 0.7$ and thus
$\sigma_v = 0.49$. This translates to an uncertainty in $\log \phi(L)$
of ${+0.2}/{-0.3}$, which is ample to account for the difference
between the luminosity functions of the two fields.

\begin{table*}
  \centering
  \begin{tabular}{lccccc}
    \hline\hline
    \centercol{Reference} & \centercol{Redshift range} & \centercol{Sky area}    & \centercol{Co-moving volume}            & \centercol{$\sigma_v$} & \centercol{$\Delta \log \phi(L)$}\\
    \centercol{}          & \centercol{}               & \centercol{(sq.~deg.)}  & \centercol{($\pow{3}$\,\perMpc{})}      & \centercol{}           & \centercol{($\phi(L)$ in \perMpc{})}\\
    \hline
    \citet{Gallego95}      & $z \leq 0.045$            & 471   & $3.3\times\pow{2}$ & 0.21 & ${+0.1}/{-0.1}$\\
    \citet{Fujita03}       & $0.233 \leq z \leq 0.251$ & 0.255 & 3.9                & 0.56 & ${+0.2}/{-0.4}$\\
    \citet{Hippelein03}    & $0.238 \leq z \leq 0.252$ & 0.086 & 1.4                & 0.70 & ${+0.2}/{-0.5}$\\
    \citet{Ly07}           & $0.233 \leq z \leq 0.251$ & 0.255 & 4.7                & 0.63 & ${+0.2}/{-0.4}$\\
    \\
    This paper, CDFS field & $0.229 \leq z \leq 0.261$ & 0.262 & 9.4                & 0.49 & ${+0.2}/{-0.3}$\\
    This paper, S11 field  & $0.229 \leq z \leq 0.261$ & 0.230 & 8.3                & 0.49 & ${+0.2}/{-0.3}$\\
    \hline\hline
  \end{tabular}
  \caption{The survey geometries for a sample of narrowband surveys
    with well-defined survey volumes alongside their root cosmic
    variance and the associated uncertainty in the number density.
    Root cosmic variance was calculated using the prescription of
    \citet{Somerville04} assuming bias $b = 0.7$ and a number
    density of line emitters of 0.05\,\perMpc{}, following
    \citet{Ly07}.}
  \label{tab:cosmicvar}
\end{table*}

Many of the narrowband surveys exhibit similar uncertainties which are
sufficiently large to account for the differences between each other.
Table~\ref{tab:cosmicvar} shows resulting uncertainty in the number
density due to the cosmic variance for a sample of narrowband surveys
with well-defined survey volumes. Despite the low redshift,
\citet{Gallego95} span a large enough volume that their uncertainty
due to cosmic variance is somewhat lower than the surveys at higher
redshift. Comparing the uncertainties $\Delta \log \phi(L)$ to the
spread of luminosity functions in
Figure~\ref{fig:schechtercomparison}, we observe that cosmic variance
is one of the dominating factors in the determination of an average
\ha{} luminosity function at these redshifts.

Finally, we make the observation that there is a high degree of
correlation between the three Schechter parameters. This is clearly
demonstrated by the confidence limit contours in the bottom panels of
Figure~\ref{fig:halumfie} and the correlation matrices in
Table~\ref{tab:schechterfit}.

\subsection{Star formation density}
\label{subsec:sfd}
The amount of extinction-corrected \ha{} luminosity from an \HII{}
region is directly proportional to the quantity of UV ionising flux
produced by newborn stars. As such, it can be used to estimate the
number of new stars and hence the star formation rate. We can thus
derive global star formation densities from the \ha{} luminosity
densities of Section~\ref{subsec:schechterfitting}. We use the star
formation rate calibration of \citet{Kennicutt98},
\begin{equation}
  \dot{\rho}~(\mathrm{M_\odot\,yr}^{-1}) = 7.9 \times \pow{-42}
  L(\mathrm{\ha{}})~(\mathrm{ergs\,s^{-1}\,cm^{-2}})~,
\end{equation}
which assumes a Salpeter initial-mass function, case B recombination
and an electron temperature of $\pow{4}$\,K.

In the case of some surveys
\citetext{\citealp{Gallego02,Hippelein03,Perez03}; this paper}, the
faint-end slope of the \ha{} luminosity function is poorly
constrained, thus having important consequences for the integrated
luminosity density. To overcome these, and in order to make a fair
comparison, we calculate the star formation density of other \ha{}
emission line surveys at the same redshift by assuming a common fixed
limit rather than integrating from zero luminosity. We choose
$\dot{\rho}_\mathrm{lim}=0.33$\,\Msunyr{}, which corresponds to the
limit of our survey ($\log F_\mathrm{lim} = -16.0$ with
$F_\mathrm{lim}$ in \lineunits{}, or $\log L_\mathrm{lim} = 40.6$ with
$L_\mathrm{lim}$ in \ergs{}), and avoids faint-end extrapolations or
assumed faint-end fits of some other surveys. This yields $\log
\dot{\rho} (L > L_\mathrm{lim}) = -2.24 ^{+0.11} _{-0.14}$ and $-1.93
^{+0.08} _{-0.10}$ for the S11 and CDFS fields, respectively.

\begin{figure}
  \centering
  \includegraphics[width=\columnwidth, trim=15pt 8pt 5pt 2pt, clip]{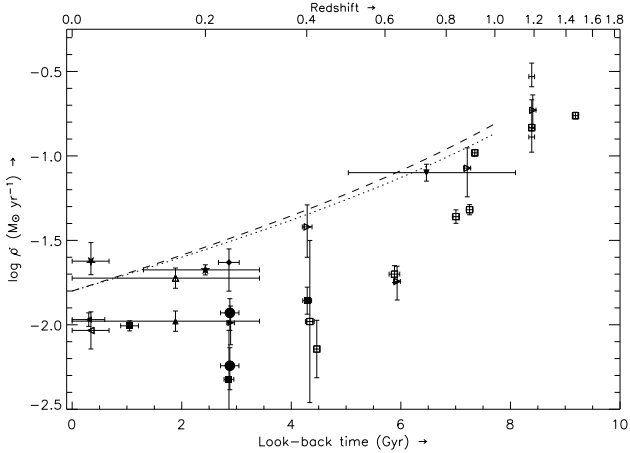}
  \caption{Star formation density as a function of look-back time
    derived from emission line surveys, where the Schechter function
    has been integrated from the star formation rate corresponding to
    the flux limit of our survey, $1\times\pow{-16}$\,\lineunits{}
    (0.33\,\Msunyr{}). The solid symbols represent the star formation
    density derived from the \ha{} line, the open symbols from either
    the \oii{} or \oiii{} line. The solid circles are the star
    formation density from the CDFS and S11 fields of this paper (top
    and bottom symbol, respectively). Other data are \citet[open and
    solid diamonds]{Fujita03}, \citet[open and solid upward-pointing
    triangle]{Sullivan00}, \citet[solid downward-pointing
    triangle]{Tresse02}, the \citet[open and solid squares]{Ly07},
    \citet[open and solid right-pointing triangles]{Hippelein03},
    \citet[solid left-pointing triangle]{Gallego95}, \citet[open
    left-pointing triangle]{Gallego02}, \citet[solid upward-pointing
    star]{Tresse98} and \citet[solid downward-pointing star]{Perez03}.
    The dotted and dashed line are the least-squares fit from
    Figures~1 and 2 of \citep{Hopkins04}, respectively. They are not
    corrected for the fact that \citet{Hopkins04} integrated the
    Schechter function down to $L = 0$\,\ergs{} and are indicated for
    comparison purposes only. The parameters used to make this Figure
    are given in Table~\ref{tab:surveypars}.}
  \label{fig:collation}
\end{figure}

Our two fields are indicated in Figure~\ref{fig:collation}. The other
results included in this Figure are derived in the same way as
described with Eq.~\eqref{eq:lumintegration} in
Section~\ref{subsec:schechterfitting}. We included only star formation
densities from surveys based on emission lines and transformed
onto the same cosmology. The majority of these points were calculated
using the compilation of \citet{Ly07}. We also included the
least-squares fit to the $z<1$ points of \citet{Hopkins04} as a point
of reference. Note that this fit assumes $L_\mathrm{lim} = 0$.

Observe that the star formation density in both our fields agrees
quite well with other \ha{} emission line surveys at the same
redshift. Nevertheless, there is a difference of almost 1\,dex between
the highest and lowest value for the star formation density. The
highest value comes from \citet{Fujita03}, which \citep[according
to][]{Ly07} suffers from contamination of higher redshift emission
line galaxies, pushing their value upwards accordingly. Observe in
Figure~\ref{fig:collation} that we have also plotted the star
formation density fits of \citet{Hopkins04} which, unlike the points,
make use of star formation density values integrated down to zero
luminosity. This serves to illustrate the extent to which
extrapolation of the faint end fit affects the final determination of star
formation density: typically up to $\lesssim50$\,\% for
$\alpha\sim-1.3$ (larger for steeper values). As discussed earlier,
the luminosity functions of several surveys have ill-constrained faint
end values.

Obviously the same systematic uncertainties discussed in
Section~\ref{subsec:comp} will also play a role here. Furthermore,
since we compare the star formation density over a larger redshift
range, other emission line star formation indicators have been used
(most notably \oii{}), thereby introducing their own sources of
systematic uncertainty In the case of \oii{}, extinction corrections
are larger and its star formation rate calibrator depends on the
abundance of the ionised gas \citep{Kewley04}. Corrections for both
can be made with spectra covering \ha{}, \hb{} and \oiii{}, as well as
\oii{}. However, at redshifts $z\gtrsim0.5$ these lines are
progressively lost from the optical, giving rise to uncertainties of
up to 0.4 in $\log (\mathrm{SFR})$, when applying the
\citet{Kennicutt92} calibrations \citep{Kewley04}. Beyond this,
emission line analyses are pushed into the near-infrared
\citep{Glazebrook99,Doherty06}, where brighter night-sky background
and instrument thermal contributions increase the difficulty of making
observations.

\subsection{Minor contaminating effects}
\label{subsec:contamination}

We now turn our attention to some additional sources of contamination
for which we have not included any correction due to their minor
nature. These are (i) the effect of the \Nii{} lines on the
measurement of \ha{} flux, (ii) the presence of active galactic nuclei
(AGNs) in the sample, and (iii) foreground \sii{} emitters.  We
consider each of these effects in turn.

\subsubsection{Narrowband \ha{} flux and \nii{}}
\label{subsec:nii}

\begin{figure}
  \centering
  \includegraphics[width=\columnwidth, trim=22pt 60pt 5pt 40pt, clip]{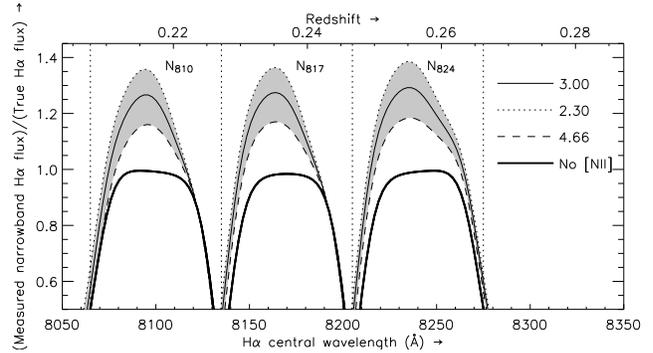}
  \caption{Measured narrowband \ha{} flux in each WFILAS filter as a
    function of redshift, expressed as a fraction of the true \ha{}
    flux. The thick solid line shows the measured \ha{} flux in the
    absence of \nii{} and approximately traces the filter transmission
    curve. The shaded envelope shows the effect of the \nii{} line
    over the range \ha{}/\nii{}$_\mathrm{tot}$ = 2.30 (dotted) to 4.66
    (dashed), centred on 3.00 (thin solid line).  The vertical dotted
    lines indicate where the filter transmission is 50\,\%.}
  \label{fig:niicontam}
\end{figure}

\begin{table}
  \centering
  \begin{tabular}{clcccc}
    \hline\hline
    $\frac{\ha{}}{\nii{}_\mathrm{tot}}$ & \centercol{Reference} & $\log \frac{\nii{}}{\ha{}}$ & \multicolumn{3}{c}{$\frac{\mathrm{\ha{} (narrowband)}}{\mathrm{\ha{} (true)}}$}\\
    & & & \na{} & \nb{} & \nc{} \\
    \hline
    2.30 & \citet{Pascual07} & $-0.49$ & 1.04 & 1.04 & 1.14 \\
    3.00 & this paper        & $-0.60$ & 0.99 & 0.99 & 1.08 \\
    4.66 & \citet{Ly07}      & $-0.79$ & 0.93 & 0.93 & 1.00 \\
    \hline\hline
  \end{tabular}
  \caption{Ratio of \ha{} to combined \nii{} line flux for a
    range of values in the literature (as indicated). The third
    column shows the corresponding values of log of the ratio of the
    \Niib{} to \ha{} line flux. The last three columns indicate the
    mean ratio of the \ha{} flux measured in the narrowband filters
    to the true flux where the filter transmission is $\ge$
    50\,\%.}
  \label{tab:niicontam}
\end{table}

The filters used in this survey are sufficiently narrow (\fwhm
$=70$\,\AA{}) that the forbidden lines of \Nii{} can affect the
measurement of \ha{} flux through narrowband imaging. The proximity of
lines complicates the estimate of \ha{} line flux in two ways. In the
case where \ha{} is central to the filter, both \nii{} lines
contribute to the narrowband-derived flux attributed to the \ha{}
flux. In the case where \ha{} is near the filter edge, then an \nii{}
line is either lost to an adjacent filter (where it contributes to the
measured continuum) or lost to the filters altogether. Due to the
finite width of emission lines and the steepness of the filters, this
is a gradual transition.

We have calculated the \ha{} flux as measured from the narrowband
imaging and compared it to its true value over a range of galaxy
redshifts, taking into account our filter setup. This was done by
taking the spectrum of one of our galaxies and fitting the \ha{} line,
as well as the \nii{}- and \sii{} doublets. We then convolved the fit
with each of our filter profiles\footnote{We have used Butterworth
  curves for the transmission profiles. The curve is given by
  $T(\lambda) = \left \{ 1 + \left ( \frac{\lambda - \lambda_c}{h/2}
    \right ) ^{2n} \right \} ^{-1}$, where $\lambda_c$ is the central
  wavelength, $h$ is the \fwhm{} of the filter and $n$ controls the
  steepness of the filter edges. We assumed $\lambda_c$ to be 8100,
  8170 and 8240\,\AA{} for \na{}, \nb{} and \nc{}, respectively, and
  \fwhm{} = 70\,\AA{} and $n$ = 3.} and calculated the line flux from
the filters in the same way as for the survey. A range of different
values of \ha{}/\nii{}$_\mathrm{tot}$ have been used, since this ratio
depends on metallicity
\citep[e.g.][]{Osterbrock89,Kewley01,Kauffmann03}. We have used two
extreme values \citetext{2.30 by \citealp{Pascual07} and 4.66 by
  \citealp{Ly07}} to reflect the wide range of metallicities found in
these galaxies. A third value of 3.00 was measured from a high quality
spectrum used for emission line fitting.

The results are shown in Figure~\ref{fig:niicontam}. Average values of
the ratio of measured to true \ha{} flux are indicated in
Table~\ref{tab:niicontam}. The narrowband \ha{} flux overestimates the
true flux by about 10\,\% when averaged over all redshifts pertaining
to a specific filter. However, the ratio can peak around 40\,\% in the
innermost 15\,\% of filter coverage. This peak corresponds to the
specific case of an idealised square filter containing all three lines
as calculated by \citet{Pascual07}. This is a worse case scenario that
only occurs rarely in practise. In the vast majority of cases the
effect of \nii{} is moderated by the sloping edges of a real filter
profile, or complete absence of \nii{} from the narrowband filters
altogether. Given the 70\,\AA{} width of our filters and the
$\sim50$\,\AA{} width of the \ha{}/\nii{} group, the chance of having
all three lines in the same filter is uncommon.

As the overall effect of \nii{} is approximately 10\,\% (corresponding
to 0.04 in $\log L$), we do not make any correction for it.

\subsubsection{AGN contribution}
\label{subsubsec:agn}

\begin{figure}
  \centering
  \includegraphics[width=\columnwidth, trim=16pt 8pt 18pt 18pt, clip]{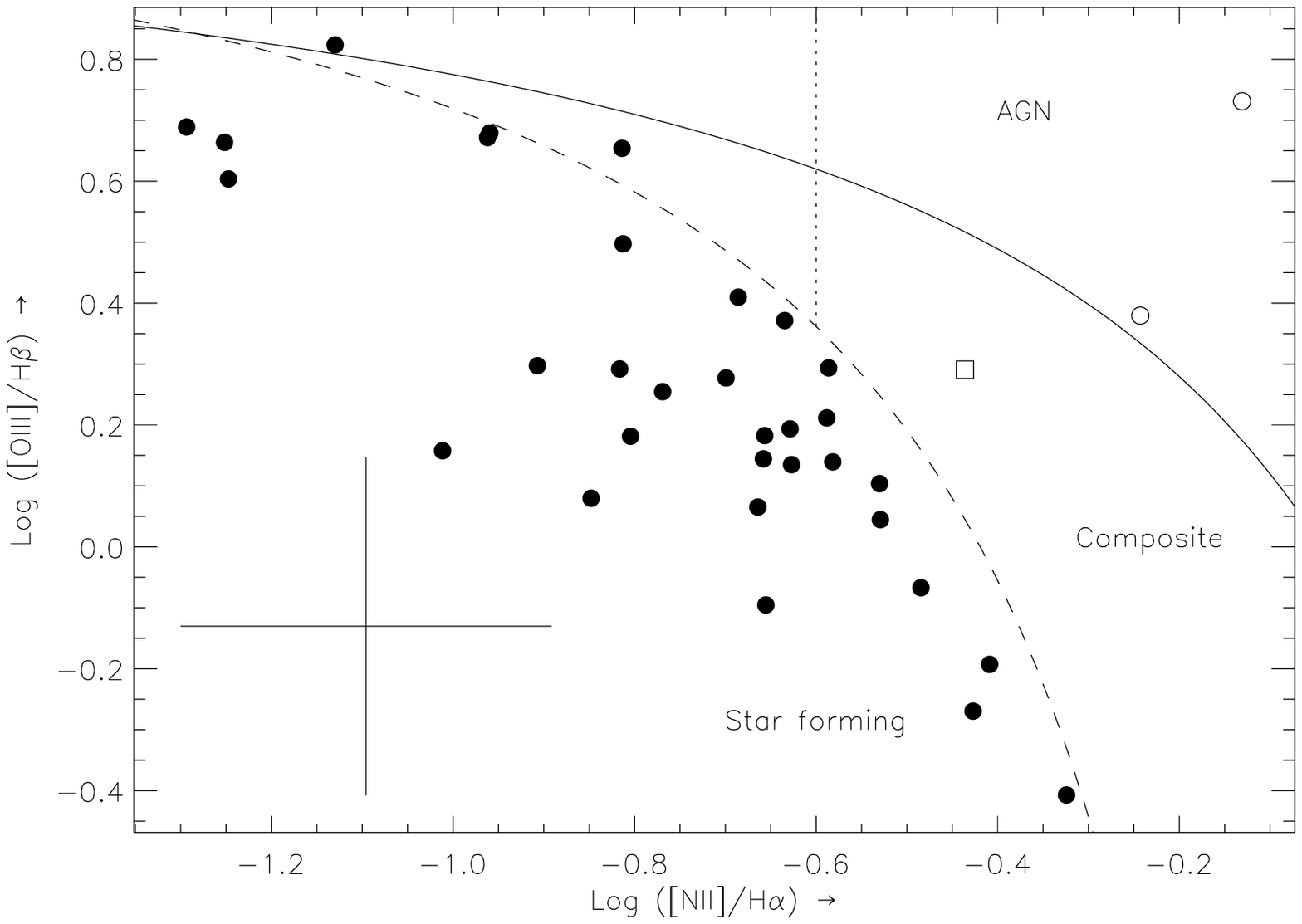}
  \caption{BPT diagram for 35 galaxies where the fluxes of the
    emission lines \hb{}, \Oiiib{} and \Niib{} have been measured with
    a signal-to-noise ratio $\ge2$. Indicated by the solid line is the
    extreme starburst demarcation of \citet{Kewley01} and the dashed
    line the demarcation of pure star formation of
    \citet{Kauffmann03}. The vertical line is drawn at $\log
    (\nii{}/\ha{}) = -0.6$. The median uncertainty is indicated by the
    error bars. The galaxies indicated by the open circles lie above
    the \citet{Kewley01} line and are most likely influenced by AGN
    activity. A third galaxy, marked by an open square, lies within
    the two demarcations and is a composite source. The remaining
    galaxies, shown by the closed circles, lie below the
    \citet{Kauffmann03} relation and left of the $\log (\nii{}/\ha{})
    = -0.6$ line and are pure star forming galaxies.}
  \label{fig:bpt}
\end{figure}

The presence of an active nucleus in a galaxy can contribute \ha{}
line flux in addition to that due to normal star formation.  For
example, \citet{Pascual01} have found approximately 15\,\% of their
luminosity density to be due to galaxies identified as AGNs.  We
computed the fraction of \ha{} contribution due to AGN in our sample
using the line diagnostic relations as determined by \citet{Kewley01}
and \citet{Kauffmann03}. We selected galaxies from both fields where
the fluxes of the emission lines \hb{}, \Oiiib{} and \Niib{} have been
measured with a signal-to-noise ratio $\ge2$.  The line ratios and the
line diagnostic relations are indicated in the
Baldwin-Phillips-Terlevich \citep[BPT;][]{Baldwin81} diagram of
Figure~\ref{fig:bpt}.

Two of the galaxies lie above the extreme starburst demarcation of
\citet[solid line]{Kewley01} and have $\log (\nii{}/\ha{}) \ge -0.6$,
which classifies them as AGNs. A third galaxy, also with $\log
(\nii{}/\ha{}) \ge -0.6$, lies below this demarcation, but above the
pure star formation boundary of \citet[dashed line]{Kauffmann03},
making this galaxy a composite case.

If we consider the total \ha{} contribution from the two AGNs in it
amounts to 5\,\% of the total \ha{} flux from this subsample.
Overall, this AGN contribution would result in a decrease in the star
formation density of $\log \dot{\rho} = 0.02$. If we include the
composite galaxy as a third AGN, the decrease is $\log \dot{\rho} =
0.04$.

\subsubsection{\sii{} contribution}
\label{subsubsec:siicontam}

\begin{table*}
  \centering
  \begin{tabular}{ccccccc}
    \hline\hline
    \multicolumn{3}{c}{CDFS} & & \multicolumn{3}{c}{S11}\\
    \cline{1-3}\cline{5-7}
    $\alpha$ & $\log L^*$ & $\log \phi^*$ & &
    $\alpha$ & $\log L^*$ & $\log \phi^*$\\
    $-$1.01 $\pm$ 0.32 & 41.28 $\pm$ 0.15 & $-$1.88 $\pm$ 0.19& &
    $-$1.17 $\pm$ 0.50 & 41.23 $\pm$ 0.24 & $-$2.24 $\pm$ 0.34\\
    \cline{1-3}\cline{5-7}
    \multicolumn{3}{c}{
      $ \left ( \begin{array}{rrr}
           1.00000 & -0.91849 &  0.95680 \\
          -0.91849 &  1.00000 & -0.96615 \\
           0.95680 & -0.96615 &  1.00000 
         \end{array} \right ) $
    } & &
    \multicolumn{3}{c}{
      $ \left ( \begin{array}{rrr}
           1.00000 & -0.90882 &  0.95080 \\
          -0.90882 &  1.00000 & -0.97085 \\
           0.95080 & -0.97085 &  1.00000 
        \end{array} \right ) $
    }\\
    \hline\hline
  \end{tabular}
  \caption{Schechter parameters for the \ha{} luminosity
    functions where the \sii{} galaxies at $z\sim0.21$ are assumed
    to be \ha{} galaxies at $z\sim0.24$. They were determined using
    a Levenberg-Marquardt $\chi^2$ minimisation. The correlation
    matrices $\rho_{ij}$ for each are shown below.}
  \label{tab:schechterfitsii}
\end{table*}

\begin{figure}
  \centering
  \includegraphics[width=\columnwidth, trim=13pt 47pt 18pt 40pt, clip]{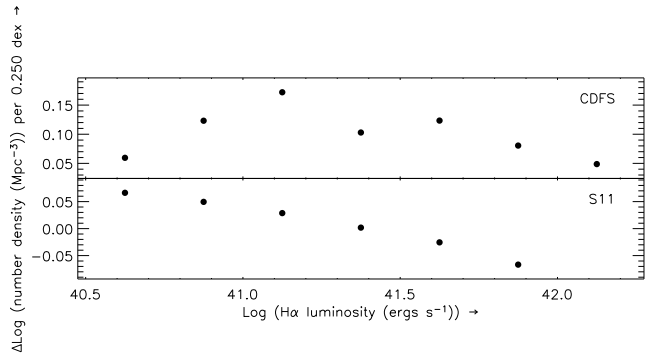}
  \caption{Differences between our derived \ha{} luminosity
    distribution (Figure~\ref{fig:halumfie}) and that assuming all
    \sii{} galaxies at $z\sim0.21$ to be \ha{}. The values for the
    Schechter parameters to the luminosity diagram including the
    \sii{} galaxies are indicated in Table~\ref{tab:schechterfitsii}.}
  \label{fig:lfs_sii}
\end{figure}

Surveys that only apply colour criteria to determine the nature of the
emission line are unable to distinguish between \ha{} emitters at
$z\sim0.24$ and \sii{} emitters at $z\sim0.21$
(Section~\ref{subsec:spectroscopy} and
Figure~\ref{fig:BRcol_histo_elg}). The previous calculations of the
\ha{} fraction assume that it is possible to distinguish between \ha{}
and \sii{} emitters. Since this can only be done with spectroscopy, we
have also derived the \ha{} luminosity function for the case where
\sii{} emitters are taken to be \ha{} emitters. This gives an
indication of the impact of having sole reliance on colour criteria
and no spectroscopic follow up. Figure~\ref{fig:lfs_sii} shows the
difference between \ha{} luminosity functions for the CDFS and S11
fields where the \sii{} galaxies were assumed to be \ha{}.  The
Schechter parameters that belong to the alternative Schechter
functions are given in Table~\ref{tab:schechterfitsii}.

The star formation density as determined from these hypothetical
\sii{}-as-\ha{} luminosity functions down to our survey limit are
$\log \dot{\rho} (L > L_\mathrm{lim}) = -2.22 ^{+0.10} _{-0.13}$ and
$-1.81 ^{+0.070} _{-0.083}$ for the S11 and CDFS fields, respectively.
In the S11 field there were only a handful of galaxies (4) identified
at $z\sim0.21$, while the CDFS contained a significant number (30).
Hence the results of the CDFS field are more significantly affected
and an increase in the star formation density $\log \dot{\rho}$ in
this field of 0.12 (about 30\,\%) can be seen. This exercise
demonstrates how a foreground overdensity of \sii{} emitters, if
present, can significantly influence the \ha{} star formation density.
For this reason, it is imperative to have at least some spectroscopic
follow up to a narrowband survey to make such situations obvious.

\section{Environmental properties}
\label{sec:environment}

The suppression of star formation rates at the centres of clusters has
been well established both through direct observation
\citep{Lewis02b,Balogh97,Balogh98,Kodama01,Kodama04,Gomez03}, as well
as a changing mix of morphological types \citep{Dressler80}. Such high
density environments provide a range of dynamical mechanisms whereby
galaxy encounters rapidly strip gas from any potential star forming
galaxies \citep[e.g.][and references therein]{Couch01}. Recent
observations have suggested a continuation of this trend across
structures at larger scales and lower density enhancements than
clusters \citep{Gomez03,Gray04}. Accordingly we examine our two fields
for evidence of star formation rates that are driven by either the
general galaxy environment, or alternatively, the local distribution
of star forming galaxies.

Usually, the amount of galaxy clustering is expressed as a function of
projected density
\begin{equation}
  \Sigma_n = \frac{n}{\pi r_n ^2}~,
\end{equation}
where $r_n$ (in Mpc) is the distance to the $n$th (usually $n=10$)
nearest neighbouring galaxy with $M_B < -19$. In cluster environments
the star formation rate has been observed to be quenched at galaxy
densities above 1\,\perMpcSq{} \citep{Lewis02b,Gomez03}.

\begin{figure}
  \centering
  \includegraphics[width=\columnwidth, trim=20pt 8pt 5pt 13pt, clip]{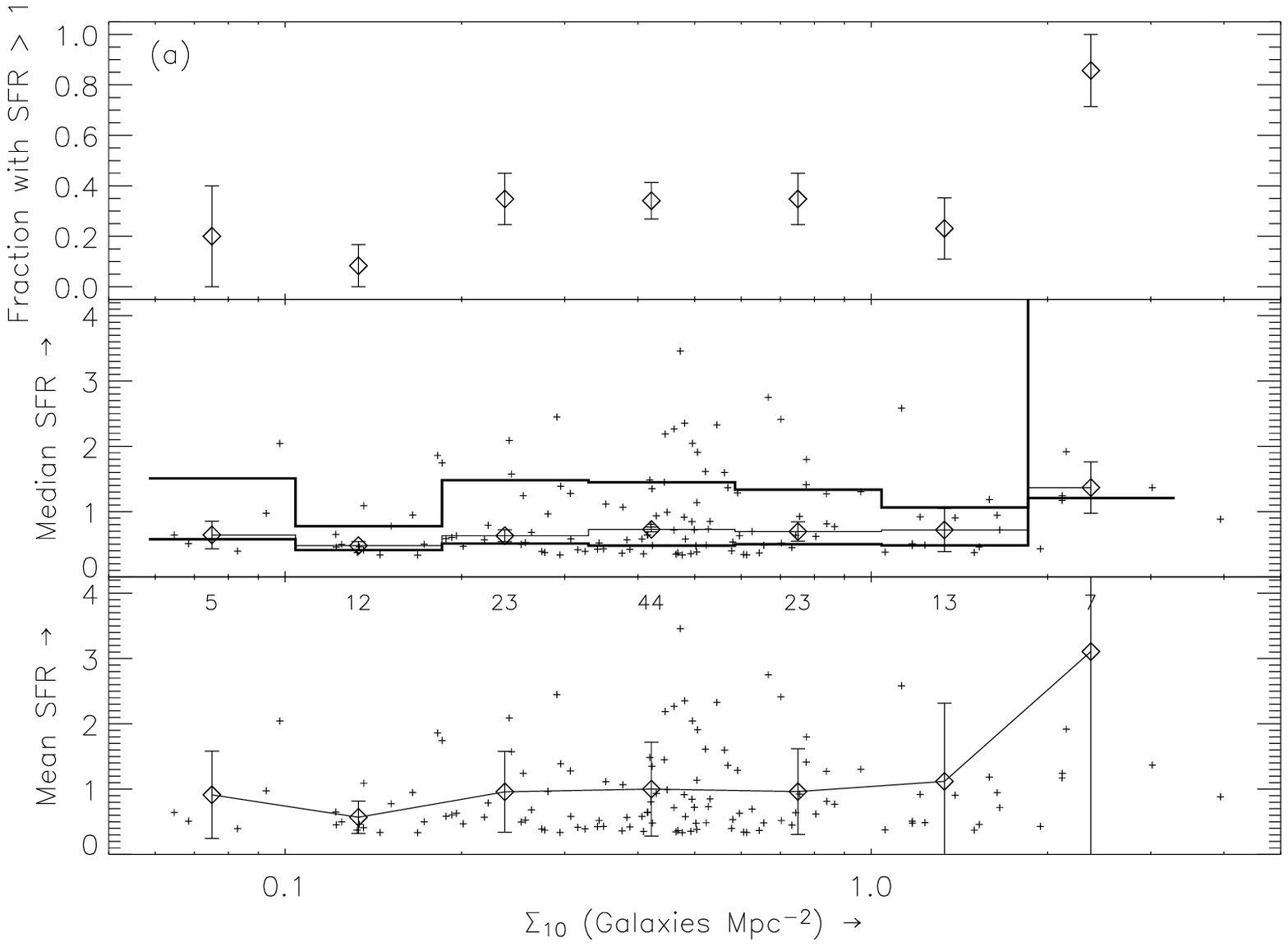}
  \includegraphics[width=\columnwidth, trim=20pt 8pt 5pt 13pt, clip]{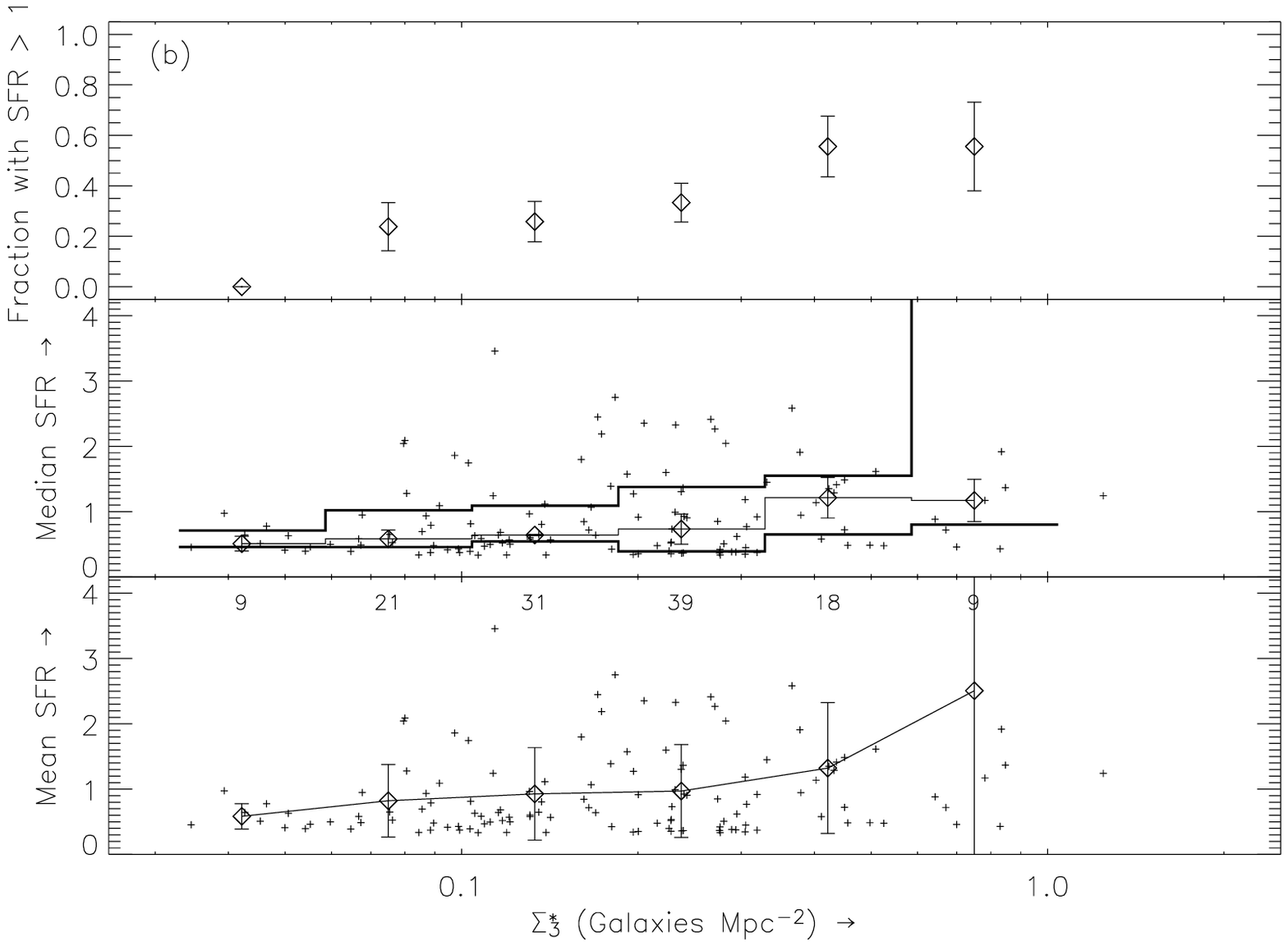}
  \caption{({\it a}) The mean and median star formation rate per
    galaxy (in \Msunyr{}) and the fraction of galaxies with a star
    formation rate $> 1$\,\Msunyr{} as a function of the projected
    density $\Sigma _{10}$ of ordinary galaxies (taken from the
    COMBO-17 survey). The errorbars in the two top panels are the
    jackknife estimates of the standard error, while in the bottom
    panel they are standard deviations. The small crosses indicate the
    values for each individual star forming galaxy. The thick lines in
    the middle panel indicate the 25th and 75th percentile for each
    bin. The numbers in the bottom panel indicate the number of
    galaxies included in each point. Some of the individual galaxies
    have values outside the range of star formation rates plotted and
    hence are not indicated. ({\it b}) Same as ({\it a}), but as a
    function of star forming galaxy density $\Sigma ^* _3$.}
  \label{fig:lewis}
\end{figure}

In Figure~\ref{fig:lewis}(a) we show the fraction of galaxies with a
star formation rate exceeding 1\,\Msunyr{}, as well as median and mean
star formation rate per galaxy as a function of the projected density
of the general galaxy population. This uses data for all of the
spectroscopically confirmed star forming galaxies at $z\sim0.24$ for
both of our fields combined. The indicated errorbars in the two top
panels were determined using the jackknife estimator\footnote{The
  jackknife estimator is calculated as follows. Let $\hat{\rho}_{(i)}
  = \hat{\rho} (x_1, \dots, x_{i-1}, x_{i+1}, \dots, x_n)$ be the
  value of the statistic with one element $x_i$ removed, and define
  $\hat{\rho}_{(\cdot)} = (1/n) \sum ^n _{i=1} \hat{\rho}_{(i)}$.
  Then $\hat{\sigma}_J^2 = (n-1)/n \sum ^n _{i=1} (\hat{\rho}_{(i)} -
  \hat{\rho}_{(\cdot)})^2$ is the square of the jackknife estimate of
  standard error \citep{Efron83}.}, while in the bottom panel they are
the standard deviation. We also show the 25th and 75th percentile
values for each bin in the middle panel. We determined the projected
density by using the usual $r_{10}$ measure of the tenth-nearest star
forming galaxy to each ordinary galaxy. Ordinary galaxies were taken
from the photometric redshift catalogues of the COMBO-17 survey
\citep[][K.~Meisenheimer, priv. comm.]{Wolf03} as galaxies with
$B_{AB} < 22$ (corresponding to $M_B = -19$) between $0.21 \leq z \leq
0.29$. As the thickness of the redshift slice influences the value of
projected density, we scale it using the difference in the thickness
of the redshift slice of our survey and the average thickness of the
$3\sigma$ cluster volumes (where $\sigma$ is the velocity dispersion
of the cluster) used in \citet{Lewis02b}.

Since we did not target any known clusters with our fields, we expect
that there will be little or no evidence for star formation
suppression in our fields. Typically, the projected density for
galaxies within the virial radius of a cluster is $\sim4$\,\perMpcSq{}
and at the centre of some rich clusters can be as high as
$10$\,\perMpcSq{} \citep{Lewis02b}. Indeed, as
Figure~\ref{fig:lewis}(a) shows, there is negligible change in the
star formation rate per unit density for the galaxies in both our
fields (noting that the highest density point is affected by poor
number statistics). Furthermore, we confirm levels of star formation
that are typical for the range of typical field galaxy densities
probed by our data as found by previous surveys
\citep[e.g.][]{Lewis02b,Gomez03}. Generally, the distribution of star
formation rates in a given density bin is rather asymmetric, making
the median a more reliable measure than the mean.

In Figure~\ref{fig:lewis}(b) we show the same measures as for (a), but
as a function of projected density of the spectroscopically confirmed
star forming galaxies at $0.23 \leq z \leq 0.26$. There are roughly
one-third as many star forming galaxies as not, and so we redefine the
projected density in terms of distance to the third-nearest galaxy,
$\Sigma ^* _3$. As a consequence, $\Sigma ^* _3$ and $\Sigma _{10}$
span a similar range of density values. We observe in
Figure~\ref{fig:lewis}(b) that star formation per galaxy increases
with increasing density. Noting again that the highest density bin is
affected by poor number statistics. Although not conclusive, this is
consistent with galaxy evolution scenarios that see galaxy-galaxy
interactions as triggers for bursts of star formation
\citep{Alonso04,Perez06}.

\begin{figure}
  \centering
  \includegraphics[width=\columnwidth, trim=0pt 8pt 85pt 18pt, clip]{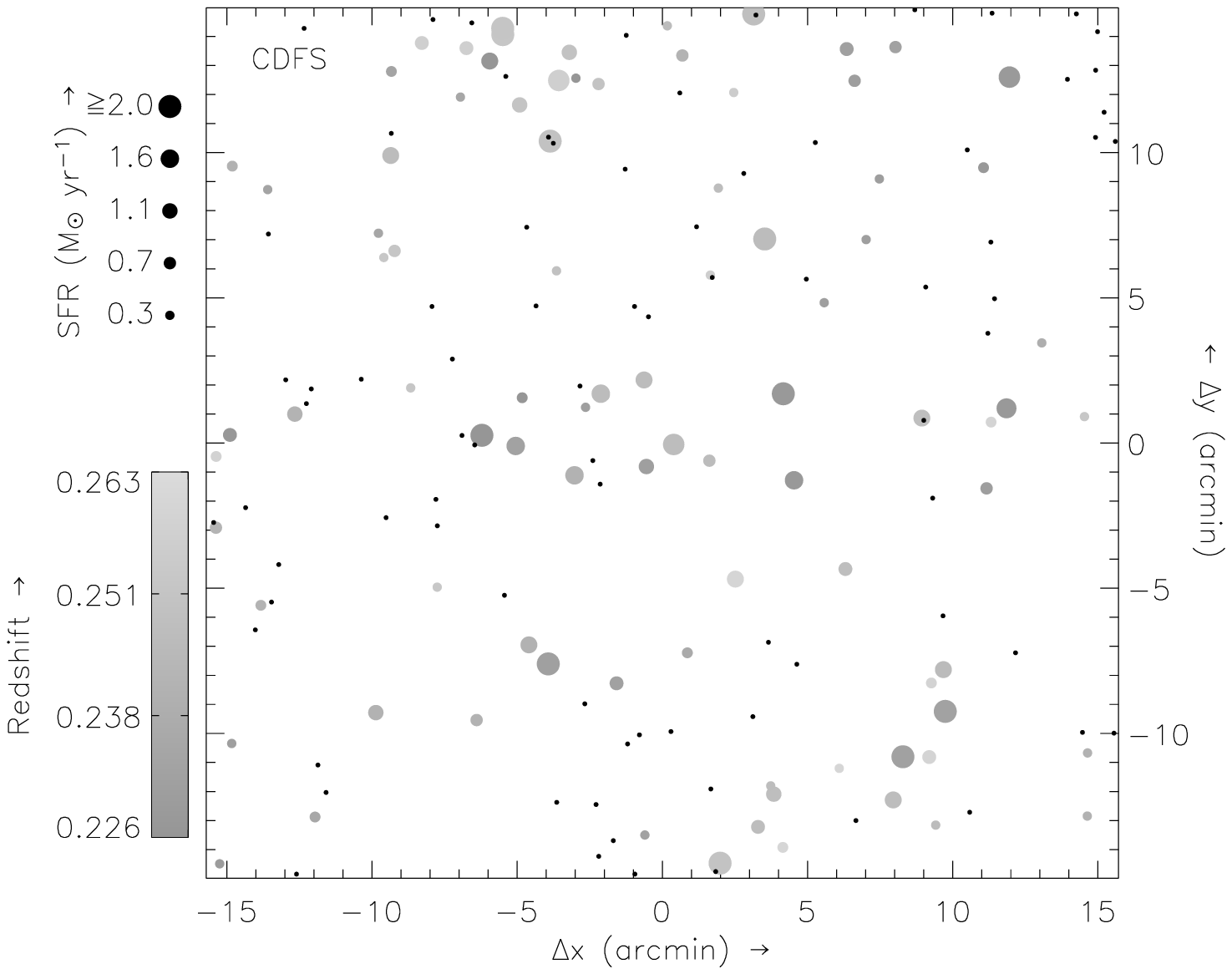}
  \includegraphics[width=\columnwidth, trim=0pt 8pt 92pt 18pt, clip]{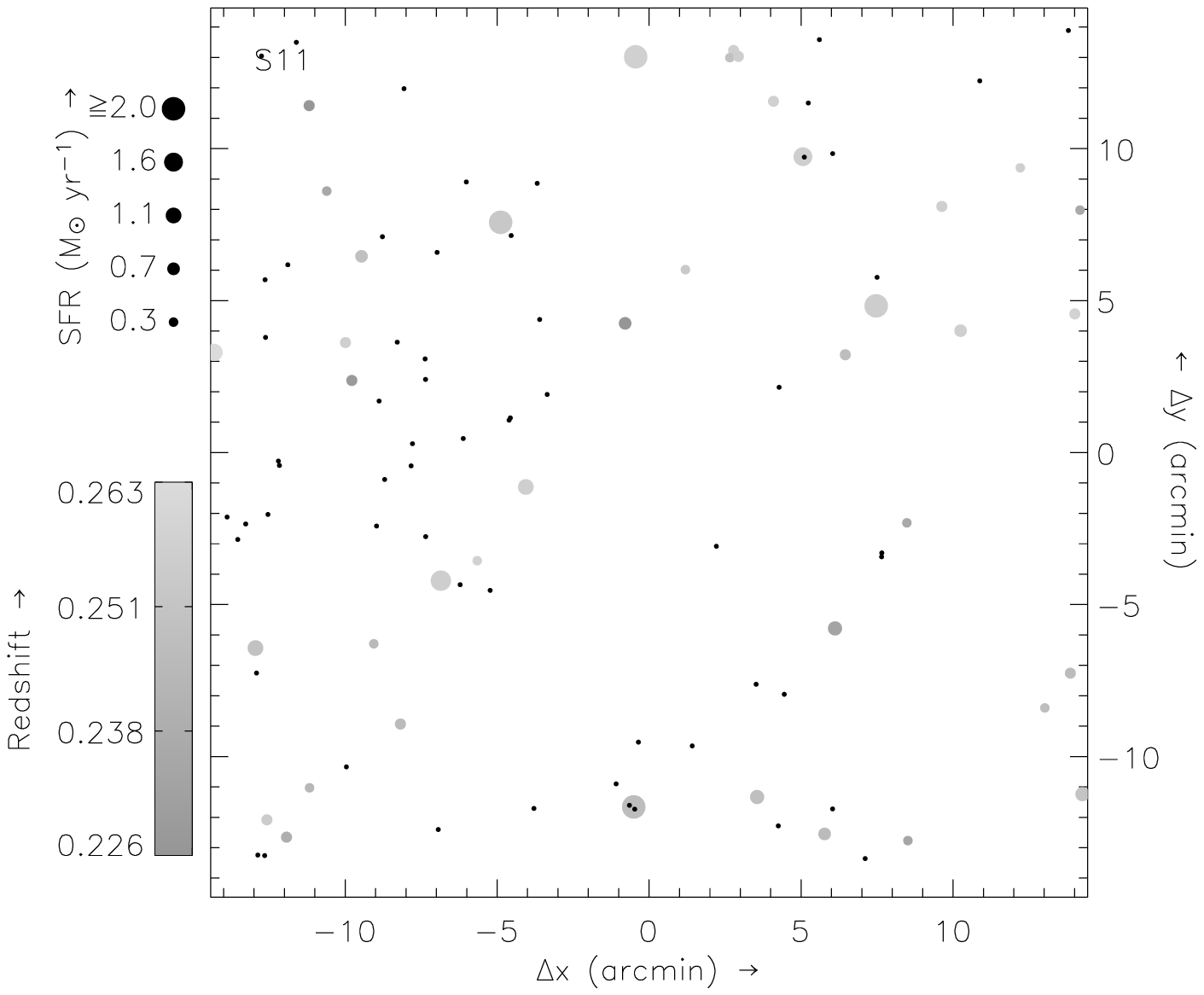}
  \caption{Spatial distribution of the spectrally confirmed \ha{}
    galaxies in both our fields ({\it solid circles}). The size of the
    circles indicates the star formation rate of the galaxy derived
    from the narrowband flux and the grey-scale the redshift. The
    black dots are galaxies that have not been spectroscopically
    confirmed yet and have a colour $0.5 \leq \BRcol{} \leq 1.3$,
    which corresponds to the colour interval of our confirmed \ha{}
    galaxies (Figure~\ref{fig:BRcol_histo_elg}).}
  \label{fig:haSkyDist}
\end{figure}

To examine the apparent relationship between star formation rate and
projected density of star forming galaxies, we plot the spatial
distribution of our spectrally confirmed \ha{} galaxies in
Figure~\ref{fig:haSkyDist}. The size of the points indicates their
star formation rate and their shade of grey the redshift. Probable
(but unconfirmed) \ha{} candidates are also shown. These were selected
on the basis of colour ($0.5 \leq \BRcol{} \leq 1.3$; see
Figure~\ref{fig:BRcol_histo_elg}) and having either indeterminate or
non-existent spectra.

The distribution of star forming galaxies in the CDFS field
(Figure~\ref{fig:lewis}b) suggests a tendency for grouping of the star
forming galaxies. However, the eye is remarkably good at making out
patterns in noisy distributions and thus we should be cautious in
these interpretations \citep[e.g. p35 of][]{Peebles93}. On the other
hand, the distribution of star forming galaxies at $z\sim0.24$ in the
S11 field (Figure~\ref{fig:haSkyDist}) is apparently less structured
than the CDFS. Because of this, we infer that the trend of increasing
star formation with rising density of star forming galaxies is largely
attributable to the data from the CDFS field. This contrast between
the fields can also be seen in differences in the \ha{}
space-densities given by the two luminosity functions in
Figure~\ref{fig:halumfie}. As a consequence, the star formation
density of the S11 field is lower than that of the CDFS
(Figure~\ref{fig:schechtercomparison}). This is due to the lower \ha{}
fraction in the S11 field compared to the CDFS field
(Figure~\ref{fig:haFrac}), to the extend that can be seen given the
more limited spectroscopy on the former.

A more robust approach would be the derivation of two-point
correlation statistics of the star forming galaxies, which could
directly test for clustering tendencies in the CDFS field compared to
S11. Such analyses are beyond the scope of this paper, but will be
addressed in a future work.

\section{Summary and conclusions}
\label{sec:conclusionsLowz}

In this paper we report the results of a survey for \ha{} emitting
galaxies at $z\sim0.24$. We used two fields from the Wide Field Imager
Lyman Alpha Search (WFILAS). It consists of imaging in three
narrowband filters (\fwhm{} $=70$\,\AA{}), an encompassing
intermediate band filter (\fwhm{} $=220$\,\AA{}), supplemented with
broadband $B$ and $R$. The narrowband filters cover a redshift range
of $0.23 \lesssim z \lesssim 0.26$ for \ha{} galaxies. These galaxies
were selected by having an excess flux in one of the narrowband over
to the other two, while also being detected in the intermediate and
broadband $R$ filters. This yielded a total of 707 candidate emission
line galaxies (after the removal of stellar contaminants) for both
fields.

Of the 372 and 335 candidates, we observed 301 and 255 through
spectroscopic follow-up for the CDFS and S11 fields, respectively.  We
have identified emission in 189 and 117 candidates and confirmed that
around half of these galaxies are \ha{} at $z\sim0.24$. A significant
number of galaxies were also found at $z\sim0.21$ by means of their
\sii{} emission. Other galaxies found were \oii{} and \hb{}/\oiii{}
emitters at $z\sim1.2$ and $z\sim0.6-0.7$, respectively.  Through use
of the spectroscopy, we refined our colour selection to account for
galaxies with a single emission line, leading to a measure of the
fraction of \ha{} galaxies as a function of narrowband flux in both of
these regions of the sky. We also used the spectroscopy to determine a
generic extinction correction using the Balmer decrement.

We have determined the \ha{} luminosity function at $z\sim0.24$
separately for both of our fields after correcting for imaging and
spectroscopic incompleteness, extinction and contamination from
interlopers. We find small differences in their slope and turn-over
luminosity while their normalisations were the same. When compared to
recent \ha{} surveys, there is remarkable agreement between the
luminosity function of our CDFS field with that one the Fabry-Perot
imaging survey of \citet{Hippelein03}. Differences between our fields
were of the order expected by cosmic variance but less than the
scatter between the \ha{} luminosity functions of recent surveys. We
surmise that while cosmic variance is a major contributor to this
scatter, it is differences in methodology between surveys (mainly
differences in selection criteria) that dominate discrepancies between
\ha{} luminosity functions and its related observables at $z\sim0.24$.
A survey that covers $10-20\times$ the volume of one of our fields is
required to get the uncertainty due to cosmic variance to the levels
of \citet{Gallego95}.

We estimated the star formation density for both our fields to be
$\log \dot{\rho} = -1.93 ^{+0.08} _{-0.10}$ and $-2.24 ^{+0.11} _{-
  0.14}$ ($\dot{\rho}$ in \Msunyr{}) for the CDFS and S11 fields,
respectively, down to our survey limit of $\log F_\mathrm{line} =
-16.0$ ($F_\mathrm{line}$ in \lineunits{}) or $\log L_\mathrm{line} =
40.6$ ($L_\mathrm{line}$ in \ergs{}). These values are comparable to
other surveys at this redshift when calculated to the same flux limit.
Correcting for AGN would decrease these values by 0.02 to 0.04
depending on exactly how much of the \ha{} flux is contributed by the
active nucleus rather than by normal star formation.

Furthermore, we determined the star formation density in the
hypothetical case where \sii{} emitters at $z\sim0.21$ were classified
as \ha{} to illustrate the problems associated with solely relying
colour selections. The star formation density $\log \dot{\rho}$ of the
S11 field does not change by much (+0.02). On the other hand, the star
formation density in the CDFS increases by 0.12, due to the large
number of foreground \sii{} galaxies at $z\sim0.21$.

We explored the amount of star formation with respect to the local
environment and found that the star formation rates were typical for
the field galaxy densities probed, in agreement with the results of
previous work. However, we also found tentative evidence of an
increase in star formation rate per galaxy with increasing density of
the star forming galaxies. This supports scenarios where merger events
are triggers for enhanced star formation, provided it can be
demonstrated to be occurring on the smallest scales. We explored this
trend by examining the spatial distribution of our fields individually
and found that it was largely attributable to one field. A formal
study of the clustering statistics of this field is required to
confirm this and will be the subject of a future study.

\section*{Acknowledgements}
We are indebted to Rob Sharp for his suggestions on preparing and
reducing the AAOmega observations and for his help during the
observations. We are also grateful to AAO service observers Will
Saunders and Quentin Parker. We thank Klaus Meisenheimer for providing
us with the photometric redshift catalogue of the S11 field. E.W. is
grateful to Philip Lah for help with Schechter function fitting. This
research was made possible by using European Southern Observatory and
Anglo-Australian Observatory facilities. We appreciate the
constructive report of an anonymous referee, which has helped to
improve the paper significantly.

\bibliographystyle{mn2e}
\bibliography{lowz.bib,astroph.bib}

\end{document}